\PassOptionsToPackage{unicode}{hyperref}
\PassOptionsToPackage{hyphens}{url}
\PassOptionsToPackage{dvipsnames,svgnames,x11names}{xcolor}
\documentclass[
  12pt,
]{article}

\usepackage{comment}
\usepackage{tabularx} 

\usepackage{natbib}
\setcitestyle{authoryear,open={(},close={)},aysep={}}
\usepackage{amsmath,amssymb}
\usepackage{setspace}
\usepackage{iftex}
\ifPDFTeX
  \usepackage[T1]{fontenc}
  \usepackage[utf8]{inputenc}
  \usepackage{textcomp} 
\else 
  \usepackage{unicode-math} 
  \defaultfontfeatures{Scale=MatchLowercase}
  \defaultfontfeatures[\rmfamily]{Ligatures=TeX,Scale=1}
\fi

\usepackage[]{mathpazo}

\usepackage{wrapfig}

\ifPDFTeX\else
\fi
\IfFileExists{upquote.sty}{\usepackage{upquote}}{}
\IfFileExists{microtype.sty}{
  \usepackage[]{microtype}
  \UseMicrotypeSet[protrusion]{basicmath} 
}{}
\makeatletter
\@ifundefined{KOMAClassName}{
  \IfFileExists{parskip.sty}{%
    \usepackage{parskip}
  }{
    \setlength{\parindent}{0pt}
    \setlength{\parskip}{6pt plus 2pt minus 1pt}}}{
  \KOMAoptions{parskip=half}}
\makeatother
\usepackage{xcolor}
\usepackage[margin=1in]{geometry}
\usepackage{longtable,booktabs,array}
\usepackage{calc} 
\usepackage{etoolbox}
\makeatletter
\patchcmd\longtable{\par}{\if@noskipsec\mbox{}\fi\par}{}{}
\makeatother
\IfFileExists{footnotehyper.sty}{\usepackage{footnotehyper}}{\usepackage{footnote}}
\makesavenoteenv{longtable}
\usepackage{graphicx}
\makeatletter
\def\maxwidth{\ifdim\Gin@nat@width>\linewidth\linewidth\else\Gin@nat@width\fi}
\def\maxheight{\ifdim\Gin@nat@height>\textheight\textheight\else\Gin@nat@height\fi}
\makeatother
\setkeys{Gin}{width=\maxwidth,height=\maxheight,keepaspectratio}
\makeatletter
\def\fps@figure{htbp}
\makeatother
\providecommand{\tightlist}{%
  \setlength{\itemsep}{0pt}\setlength{\parskip}{0pt}}
\newlength{\cslhangindent}
\newlength{\csllabelwidth}
\newlength{\cslentryspacingunit} 
 {
  \setlength{\parindent}{0pt}
  \ifodd #1
  \let\oldpar\par
  \def\par{\hangindent=\cslhangindent\oldpar}
  \fi
  \setlength{\parskip}{#2\cslentryspacingunit}
 }%
 {}
\usepackage{calc}

\usepackage{multirow}
\usepackage{wrapfig}
\usepackage{float}
\usepackage{colortbl}
\usepackage{pdflscape}
\usepackage{tabu}
\usepackage{threeparttable}
\usepackage{threeparttablex}
\usepackage[normalem]{ulem}
\usepackage{makecell}
\usepackage{xcolor}
\ifLuaTeX
  \usepackage{selnolig}  
\fi
\IfFileExists{bookmark.sty}{\usepackage{bookmark}}{\usepackage{hyperref}}
\IfFileExists{xurl.sty}{\usepackage{xurl}}{} 
\hypersetup{
  pdftitle={Scientific Talent Leaks Out of Funding Gaps},
  pdfauthor={Wei Yang Tham; Joseph Staudt; Elisabeth Ruth Perlman; Stephanie D. Cheng},
  colorlinks=true,
  linkcolor={blue},
  filecolor={Maroon},
  citecolor={Blue},
  urlcolor={blue},
  pdfcreator={LaTeX via pandoc}}

\title{Scientific Talent Leaks Out of Funding Gaps\footnote{This paper previously circulated under the title \emph{The Effect of Funding Delays on the Research Workforce: Evidence from Tax Records}. This paper uses data from the U.S. Census Bureau. Any views expressed are those of the authors and not those of the U.S. Census Bureau. The Census Bureau's Disclosure Review Board and Disclosure Avoidance Officers have reviewed this data product for unauthorized disclosure of confidential information and have approved the disclosure avoidance practices applied to this release. (DRB Approval Number: CBDRB-FY21-CES009-002, CBDRB-FY22-CES008-003, CBDRB-FY22-CES007-008, CBDRB‑FY23‑CES008-002, CBDRB‑FY23‑0333). We are grateful to Enrico Berkes, Valerie Bostwick, Matt Clancy, Holden Diethorn, Lorenz Ekerdt, Ina Ganguli, Donna Ginther, Cheryl Grim, Ian Hutchins, Danielle Li, Kyle Myers, Bruce Weinberg, and numerous seminar and conference participants for their invaluable feedback. We are also indebted to the NIH Principal Investigators who graciously allowed us to interview them.}}
\author{Wei Yang Tham\footnote{Harvard Business School \& Laboratory for Innovation Science at Harvard, \href{mailto:wtham@hbs.edu}{\nolinkurl{wtham@hbs.edu}}} \and Joseph Staudt\footnote{US Census Bureau, \href{mailto:joseph.staudt@census.gov}{joseph.staudt@census.gov}} \and Elisabeth Ruth Perlman\footnote{US Census Bureau, \href{mailto:elisabeth.perlman@census.gov}{elisabeth.perlman@census.gov}. Perlman would like to dedicate this paper to a soft money supported scientist, whose year-to-year funding uncertainty and grant non-renewal has impacted their own life.} \and Stephanie D. Cheng\footnote{Edgeworth Economics}}
\date{\today}

\makeatletter
\providecommand{\subtitle}[1]{
  \apptocmd{\@title}{\par {\large #1 \par}}{}{}
}
\makeatother

\begin{document}

\maketitle

\begin{abstract}

\noindent{We study how delays in NIH grant funding affect the career outcomes of research personnel. Using comprehensive earnings and tax records linked to university transaction data along with a difference-in-differences design, we find that a funding interruption of more than 30 days has a substantial effect on job placements for personnel who work in labs with a single NIH R01 research grant, including a 3 percentage point (40\%) increase in the probability of not working in the US. Incorporating information from the full 2020 Decennial Census and data on publications, we find that about half of those induced into nonemployment appear to permanently leave the US and are 90\% less likely to publish in a given year, with even larger impacts for trainees (postdocs and graduate students). Among personnel who continue to work in the US, we find that interrupted personnel earn 20\% less than their continuously-funded peers, with the largest declines concentrated among trainees and other non-faculty personnel (such as staff and undergraduates). Overall, funding delays account for about 5\% of US nonemployment in our data, indicating that they have a meaningful effect on the scientific labor force at the national level.}




\end{abstract}

\setstretch{1.5}

\newpage

\begin{quote}
``My current job started 8 years ago when my boss told me he had 6 months of guaranteed funding. I worked for him full-time for 4 years, my salary cobbled together from a half-dozen grants over that time\ldots[W]hile my skills are undoubtedly valuable to a research lab, it is incredibly difficult for someone like me to find a stable job because of the funding issues...'' - \href{https://guzey.com/how-life-sciences-actually-work/\#a-lab-techmanagers-email-about-their-experience-in-response-to-this-essay}{Anonymous lab technician/manager} \citep{guzey2019}

\end{quote}

\hypertarget{intro}{%
\section{Introduction}\label{intro}}

The heavy dependence of university research on federal funding is not accidental. In the aftermath of ``the scientist's war,'' Vannevar Bush, director of the WWII Office of Scientific Research and Development, laid out a vision for US science. Bush's 1945 report, ``Science: The Endless Frontier'' called for the creation of a federal science agency that would provide “stability of funds so that long-range programs may be undertaken.''\footnote{Guided by the importance of basic research to the war effort, Bush sought to ``strengthen'' universities, which he viewed as essential to the production of basic research because they were ``least under pressure for immediate, tangible results.'' See \cite{stephan2013endless} and \cite{gross2023ww2}.} The grant system that resulted, however, exhibits instabilities that harm individual scientists and ultimately science itself \citep{alberts2014rescuing}.


Today, federal funding routinely exposes researchers to grant uncertainty through, for example, the 1998–2003 boom-bust cycle of the National Institutes of Health (NIH) \citep{freeman2009doubling} and the contentious federal budgeting process with accompanying threats of government shutdowns.\footnote{The unpredictability of when a given year's budget will be passed is such a regular occurrence that it might even be considered a permanent feature of the scientific funding landscape. For example, the National Institute of Allergy and Infectious Diseases (NIAID), which accounted for 14\% of the NIH's budget in FY 2020, explicitly addresses this issue in an \href{https://web.archive.org/web/20220110153528/https://www.niaid.nih.gov/grants-contracts/due-dates-preparation-time-review-cycles}{online guide} to the grant application process, stating that it is ``assiduous about issuing awards using funds from the [continuing resolution].''} Moreover, the process of applying for and renewing grants can itself be lengthy and unpredictable, ranging between 8 and 20 months for NIH grants \citep{fikes2018, drugmonkey2009never, drugmonkey2016never, mervis1996shutdown}.\footnote{The \href{http://web.archive.org/web/20230316114209/https://www.niaid.nih.gov/grants-contracts/timelines-illustrated}{National Institute of Allergy and Infectious Diseases (NIAID) guide to grant timelines} says that ``[i]f your application succeeds on the first try, it typically takes between 8 and 20 months after the due date to get an award’'. Specifically, 8--20 months refers to the time between application submission to the arrival of funds.} 

In addition, the COVID-19 pandemic highlighted that funding delays are more than a bureaucratic nuisance; they can hinder scientific progress when it is urgently needed \citep{williams2023delays, collison2021fastgrants}. However, we lack systematic evidence on how delays might affect the careers of people supported by grant funding, which is crucial for policy because these research personnel have acquired scarce human capital, at both high personal and social cost, and play a critical role in advancing knowledge and technology. Furthermore, the impacts of funding delays today may feed into the expectations of potential scientific workers and deter talent from entering the research workforce. 

The impacts of funding delays on the scientific workforce are difficult to study because they are often caused by aggregate shocks (i.e., ~all researchers are affected by the federal budgeting process). We address this challenge by isolating lab-level variation in funding delays arising from a particular institutional feature of the NIH's most common research funding mechanism, the R01 grant. R01s are usually granted for four to five years, after which a Principal Investigator (PI) can apply for renewal after each term ends. \footnote{Receiving an R01 is generally regarded as necessary for establishing an independent research lab in the biomedical sciences \citep{new_science_nih}.} However, even if the R01 is \emph{successfully renewed}, there can be delays in the disbursement of its funds. 
Confining our analysis to \emph{successfully renewed} R01s, we measure funding delays using publicly available administrative data on NIH grants (ExPORTER), identifying treated labs that experience a delay (``interrupted'' labs) and compare them to similar control labs whose funding was not delayed (``continuously-funded'' labs).\footnote{Renewal is not guaranteed – the \href{https://report.nih.gov/nihdatabook/report/29}{success rate} of renewal applications in a given year is about 20\%.} Using a delay of more than 30 days as a baseline definition, over 20\% of these successfully renewed grants (between 2005-2018) experience an interruption.\footnote{Our choice of 30 calendar days is meant to approximate a month -- grants are usually funded on the first of the month, so the arrival of new grant funding can be thought of as occurring on a monthly basis. We also present results separately for ``short'' (30 to 90 days) and ``long'' (90 days or more) delays, which confirm that our results are not sensitive with respect to the exact definition of an interruption.}



We next link our sample of successfully renewed R01s to university administrative data on grant transactions (UMETRICS), allowing us to observe the individual personnel (e.g.,~faculty, postdocs, grad students, etc.)~supported by these grants. These personnel are then linked to their career outcomes using the universe of confidential W-2 and 1040 Schedule C (ILBD) tax records as well as unemployment insurance earnings records (LEHD), which provide their complete earnings and employment history in the US for 2005-2018. We also link these lab personnel to a variety of additional data, including PubMed publications, comprehensive administrative data on demographic characteristics (age, gender, race, and ethnicity) as well as the full 2000, 2010, and 2020 Decennial Censuses.

Using a difference-in-differences design, we compare the career outcomes of personnel in labs with an interrupted R01 to those in labs with only continuously-funded R01s. Our estimation procedure combines ``stacking'' by cohorts of grants set to expire (but ultimately successfully renewed) in a given year \citep{baker2022dd, cengiz2019minwage} and the estimator from \cite{callaway2020dd}.
We first examine how funding interruptions affect the job placements of lab personnel. After an interruption, personnel in labs supported by a single R01 are immediately 3 percentage points (pp) more likely to become nonemployed in the US (i.e.,~they do not appear in our comprehensive tax and earnings data), an almost 40\% increase.\footnote{The term ``nonemployed'' has been used before in the economics literature (e.g., \cite{murphy1997unemployment} and \cite{hornstein2014measuring}) to draw a distinction from the more common term ``unemployed'', which refers to people who are not working and actively looking for a job. We do not observe whether nonemployed personnel are actively looking for a job, and thus cannot observe whether they are unemployed.}$^,$\footnote{Section \ref{benchmark-main} describes how we use mean changes in outcomes for continuously-funded personnel to convert our estimates into percentage increases.} This US nonemployment effect persists for at least five years and, for context, is about one-third of the motherhood nonemployment effect for Ph.D.s in the biological sciences \citep{cheng2021childcare}. There is a corresponding decrease in employment probability in US industry, government, or the non-profit sector (hereafter referred to as ``industry'').

These employment effects are concentrated among trainees (graduate students and postdocs) and the US-born, who are 6.1~pp and 3.5~pp more likely to enter US nonemployment, a 60\% increase for both subsamples. For both groups, these changes are almost entirely driven by departures from universities. In contrast, we do not find an effect for any job placement outcomes of faculty and estimate that foreign-born personnel are half as likely as their US-born counterparts to be induced into US nonemployement. Thus, long term and university supported contracts appear to insulate faculty from the consequences of interruptions, while the temporary status of trainees ensures they bear the brunt. Meanwhile, the relative attachment of foreign-born personnel to university employment (and to their original university) may reflect less flexibility in altering career plans due to visa constraints or a stronger preference for staying in the US \citep{ganguli2019postdocs}.

The seemingly permanent increase, for interrupted single-R01 personnel, in the likelihood of US nonemployment raises the question of what these highly employable individuals are doing.\footnote{The Survey of Doctorate Recipients consistently finds that the unemployment rate among surveyed individuals is about 1.5\%, several times lower than the overall unemployment rate. For 2001-2013, see \href{https://www.nsf.gov/statistics/infbrief/nsf14317/}{here}. For 2015-2021, see \href{https://www.ncses.nsf.gov/pubs/nsf23318/figure/1}{here}.} Some may remain in the US, truly not working, but others may have left the US and therefore the scope of our US-based tax data. To distinguish these groups, we compute the share of the nonemployment effect attributable to personnel who are absent from the full 2020 Decennial Census (and therefore likely to be living outside the US), finding that about half have left the US entirely.\footnote{We also examine personnel presence in the US over time using the 2000, 2010, and 2020 Decennial Censuses, which also suggests that interruptions push single-R01 personnel out of the US (Appendix Table \ref{tab:tab-censusonly}).} For trainees, the results are starker, with 70\% leaving the US, likely reflecting the mobility of a relatively young population with less attachment to the US labor market and a greater willingness to find a job abroad. Breaking out by place of birth, we find that while only 30\% of the US-born induced to nonemployment left the US, nearly all of the foreign-born left, possibly reflecting a combination of push (e.g., immigration restrictions) and pull (e.g., support networks in their home countries) factors.



Since interruptions appear to encourage the exit of scientific personnel from the US, it is natural to ask whether these personnel leave the scientific enterprise altogether or whether they continue to produce scientific output elsewhere. Though our administrative earnings and Decennial data are comprehensive, they end at the US border, so we use publications to track the scientific output of personnel across the world. Computing the share of the nonemployment increase attributable to  personnel who publish, we find that the vast majority (about 87\%) of single-R01 personnel publish less actively after an interruption. Once again, the results are most dramatic for trainees, with 96\% publishing less actively. By comparison, nearly all faculty induced to nonemployment continue to publish at similar rates after an interruption.


In contrast to single-R01 personnel, personnel in labs supported by multiple R01s experience precisely estimated zero post-interruption changes for all employment outcomes. The importance of a funding cushion accords with the intuitions of PIs we interviewed and with evidence that, until the grant is renewed, interruptions lead to a spending collapse for single-R01 labs and only modest spending decreases for multiple-R01 labs \citep{tham2023interrupted}.\footnote{See Appendix Figure \ref{fig:lab_spending}.} Since time-varying confounders likely affect all labs similarly, this cushion phenomenon for multiple-R01 labs increases the credibility that our results for single-R01 labs are, indeed, driven by funding interruptions.\footnote{This is similar to a placebo test, as \textit{ex ante} we do not know that the effects of interruptions on the multiple-R01 sample are necessarily zero. See Section \ref{alternative-control-group}.} We interpret this as evidence for an intuitive mechanism: interruptions severely constrain the NIH funding of PIs with a single R01 and other sources of funding (e.g., university-provided bridge funding) cannot compensate, leaving them unable to pay the salaries of their lab personnel.

Overall, interruptions account for about 5\% of the US nonemployment among research personnel within our sample.\footnote{This is calculated by multiplying our estimated effect of an interruption on US nonemployment (3~pp) by the proportion of treated personnel (approx. 20\%) and then dividing this ratio by the proportion of all personnel who are nonemployed in the US at the end of our sample period (12.5\%). That is: $(0.03*0.2)/0.125 = 0.048$.} By comparison, a different sort of administrative delay -- green card delays -- account for about 7.4\% of departures from the US in a representative sample of US doctorates \citep{kahn2020impact}.\footnote{Appendix Section \ref{kahn2020calc} details how we calculate this number.} This indicates that policies to reduce or eliminate funding delays can have a meaningful effect on retaining scientific talent within the US. 


Losing highly-trained research personnel may be detrimental to the scientific enterprise, but it is unclear whether interruptions damage the careers of individual scientists. If interruptions push personnel into higher-paying private sector jobs, their earnings could rise relative to continuously-funded peers who remain at universities. Instead, we find that, after an interruption, the relative earnings of single-R01 personnel decline by 20\%.\footnote{Since our data do not include earnings outside the US, we obtain our main earnings estimates from a restricted sample that only contains personnel who are ``fully attached'' to the US labor market after an interruption, i.e.,~they are employed in the US every year post-interruption. A causal interpretation of these earnings estimates requires additional assumptions about post-treatment selection into the sample. We discuss these in further detail in Section \ref{wage-outcomes}.} We also find that job mobility within academia increases, though not in industry. We view the combination of an earnings decline and higher mobility as \textit{prima facie} evidence that interruptions not only prematurely push personnel out of their university, but also lead to worse job matches and job instability. As with job placement outcomes, interruptions have precisely estimated zero effects on the earnings of multiple-R01 personnel, again suggesting that these individuals are shielded from the consequences of funding delays.

In a final set of results, we probe our definition of an interruption by examining whether impacts vary by delay length. We find that longer interruptions (greater than 90 days) do not lead to stronger effects, suggesting that 
universities/PIs are unable to effectively bridge even relatively short funding gaps.\footnote{An interrupted lab is unlikely to know \textit{a priori} if or when their grant will eventually be renewed. Labs face considerable uncertainty at the time of funding expiration.}

As with any difference-in-differences approach, the plausibility of our estimates hinges on a parallel trends assumption, and there are several reasons to believe this is plausible in our setting. First, raw means and event studies suggest that interrupted and continuously-funded personnel trend similarly prior to grant expiry. Second, though not necessary for parallel trends to hold, balance statistics suggest that interrupted and continuously-funded labs are similar across a variety of pre-treatment observables, including demographic characteristics (gender, race, ethnicity, and place of birth), occupational composition, and research production. Third, as suggested, the effects of interruptions are confined to groups that, \textit{ex ante,} we would expect to be most vulnerable to funding delays -- specifically, non-faculty in labs supported by a single R01. Finally, all of our main estimates are robust to controlling for the number of resubmissions an R01 renewal application went through before approval, which is a measure of perceived quality.

To gain further insight into the causes and consequences of funding interruptions, we interviewed six PIs who have experienced these challenges. A recurring theme was constant worry about funding stability, with one PI noting that this ``can be very stressful, in general it is hard to plan research in advance, when getting most of your funding in installments and uncertainty.''  This is especially true for PIs running single-R01 labs, with one stating that ``If you have only one R01 grant, then you are really exposed to the vagaries of the funding cycle.'' The PIs agreed that when funding was tight, personnel, as the largest grant expense, were often first on the chopping block, leading to departures of lab personnel like technicians and postdocs. Notably, one PI recounted an instance where a technician, compelled to search for a new job due to funding uncertainty, had already moved on to a new position by the time the lab's funding was eventually secured.

Our work lies at the intersection of several strands of literature in labor economics and the economics of innovation. There are parallels between the funding interruptions studied in this paper and a rich literature on the impact of adverse events on the labor market \citep{oyer2006initial, oreopoulos2012recession, rothstein2021lost, huckfeldt2022understanding}. Our work differs in that it studies a unique but important labor market, complementing work on how scientific careers can be meaningfully affected by early events \citep{azoulay2021yellowberets, hill2019searching}. To the extent that research generates positive externalities, a better understanding of this market is important not only for worker welfare but also knowledge production. Moreover, acquiring the human capital necessary to enter the research workforce is individually and societally (particularly through government investments) costly and labor market outcomes today may influence the expectations of potential future researchers. 

A combination of theoretical and empirical work in the economics of innovation suggests that failure-tolerant incentive schemes which provide long-term stability can induce more risk-taking and exploration among innovators and scientists \citep{manso2011motivating, ederer2013pay, azoulay2011hhmi, myers2023money}.\footnote{Although the extent to and conditions under which this applies in science remains unclear.} In the context of science, anecdotal evidence suggests that one mechanism through which stability might lead to more innovation is that it enables scientists to make longer-term plans with respect to hiring personnel \citep{fikes2018, fagen2016mixed}. By highlighting the impact of grant instability on the entire research workforce (not just faculty or research outputs), this paper helps to build a fuller picture of the interaction between grant funding and the research workforce behind knowledge production.\footnote{See \cite{baruffaldi2021returns} for work on the role of physical capital in research production and \citet{babina2023cutting} on how funding constraints can influence the type of research outputs produced.}

Our work also relates to the literature on high-skilled immigration, particularly work studying the impact of policy on worker placement across sectors and borders (e.g. \cite{diethorn2022green}, \cite{kahn2020impact}, \cite{furtado2019settling}, \cite{amuedo2019opt}). A key difference is that we do not study effects of an immigration policy \textit{per se} (e.g.,~changes in H-1B caps, OPT extension). However, because our population of interest is mobile and has a high proportion of non-US citizens, any disruptions to the labor market naturally intersect with immigration issues. 


The rest of the paper is organized into the following sections, which the reader can jump to by clicking on the following links: \protect\hyperlink{background}{Background}, \protect\hyperlink{data}{Data}, \protect\hyperlink{estimation}{Estimation}, \protect\hyperlink{results}{Results}, \protect\hyperlink{conclusion}{Conclusion}.

\hypertarget{background}{%
\section{Background}\label{background}}

\hypertarget{nih-funding-and-r01-grants}{%
\subsection{NIH Funding and R01 Grants}\label{nih-funding-and-r01-grants}}

The NIH is responsible for an annual budget of \$30-40 billion, most of which is disbursed through research grants. The R01 is the largest grant mechanism through which the NIH funds extramural research. It is designed to provide enough funding to establish an independent research career.\footnote{In interviews with NIH Principal Investigators, all of them stated that there are no or few good substitutes for getting an R01 grant.} An R01 project period lasts for 4-5 years, after which it must be renewed in order to receive additional funding for a subsequent project period.\footnote{They can also be shorter (1-3 years), but this is uncommon.} Thus, the same \emph{project} can last for multiple \emph{project periods}.

Principal Investigators (PIs) generally want to maintain R01 funding for as long as possible, so it is expected that as their current project period ends they will apply to renew their project for another 4-5 year project period.\footnote{R01 renewal is sometimes even listed as a criterion for receiving tenure (e.g.,~\cite{osumed2020}).} In order to avoid lapses in funding between two project periods, PIs usually start to apply for renewal about a year before a project period ends, balancing the need to have made sufficient progress on their project while allowing time to prepare the renewal application itself as well as time to revise and then resubmit an application that is rejected.

The focus of our paper is on the effects of temporary interruptions, so we exclusively analyze projects that are \emph{successfully renewed} at least once and so span multiple project periods. Though all grants in our sample are eventually successfully renewed, some will experience a lapse in funding between two project periods (i.e.,~are interrupted) and others will be continously-funded. Of course, some projects are not renewed upon expiry, and these are not used in our analysis.

\hypertarget{where-do-funding-interruptions-come-from}{%
\subsection{Where do Funding Delays Come From?}\label{where-do-funding-interruptions-come-from}}

Funding delays can arise for several reasons. First, the US federal budgeting process is often fraught and rarely in place by the beginning of the fiscal year \citep{crs2023}. This introduces uncertainty into that year's NIH budget and limits funding to grants that are high priority, delaying decisions on others until there is more clarity about its budget for the fiscal year. 

\begin{figure}
{\centering \includegraphics{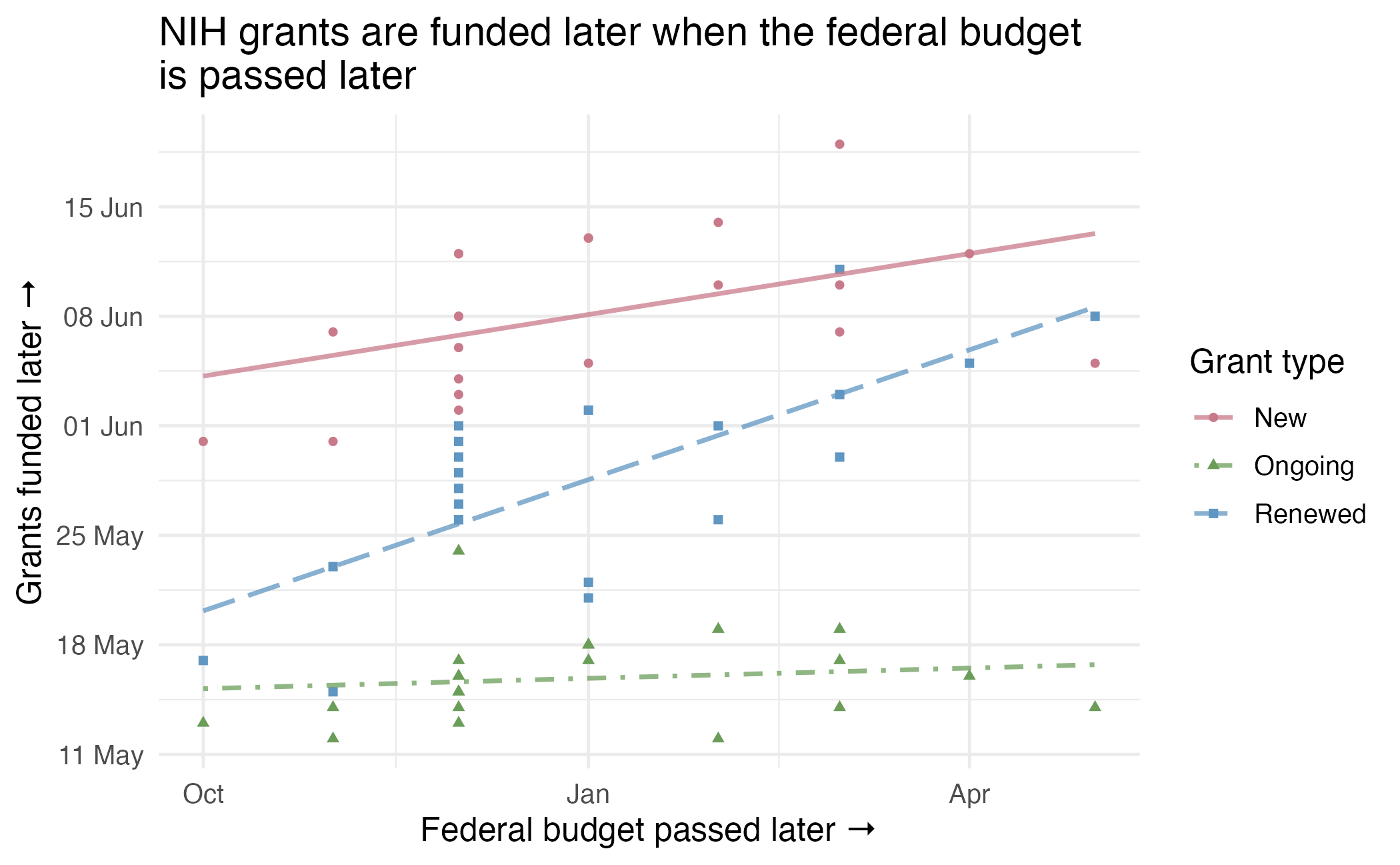} 
}
\caption{\small Each point represents a Fiscal Year (FY) from FY1998 to FY2018, and shows the average date on which NIH grants that FY were funded, plotted against the month the US federal budget was passed for that FY. ``Ongoing'' grants were already approved in previous FYs, while ``New'' and ``Renewed'' grants are newly approved in the current FY.}\label{fig:fed-v-nih}
\end{figure}

Figure \ref{fig:fed-v-nih} shows the relationship between the date in a fiscal year (1998-2018) when the federal budget was passed and the average start date of NIH grant budgets. When a grant is "Ongoing" (e.g.,~in the third year of a 5-year grant), there is no relationship. But for grants that had to be competed for (i.e.,~"New" or "Renewed" grants), budgets tend to start later in the fiscal year if the federal budget was passed later. 

Second, for any given application, the review process itself may end up being lengthy. For example, if an application is unsuccessful on its first attempt, the PI must revise and resubmit their proposal, adding time to the approval process. Indeed, we estimate that, among R01 grants that are eventually successfully renewed, each additional resubmission is associated with a 12~pp increase in the probability of being interrupted (i.e.,~there are more than 30 days between expiry and renewal, per the definition we use in our main analysis). As described later, we use resubmissions to control for (perceived) quality and isolate variation in interruption status for grants with the same number of resubmissions.

\hypertarget{why-interruptions-may-affect-employee-outcomes}{%
\subsection{Why Interruptions May Affect Employee Outcomes}\label{why-interruptions-may-affect-employee-outcomes}}

When an interruption occurs, a lab must decrease spending. Since payroll typically comprises the bulk of a grant's expenditures, the PI's ability to continue supporting personnel is circumscribed, especially if the interrupted R01 is the lab's only funding source.\footnote{In the sample of labs from \citet[Section ``Descriptive Statistics'']{tham2023interrupted}
, the median single-R01 lab spent \$13,900 on labor payments and \$900 on vendor payments in the month one year prior to R01 expiry.} Beyond funding lapses per se, even the \emph{potential} of an interruption may affect lab personnel through uncertainty over if or when funding will arrive. This uncertainty can create ambiguity surrounding the future employment prospects of personnel, possibly driving some to leave their current lab positions before funding actually runs out. For example, one PI we interviewed described the university's union contract requiring them to warn personnel six months in advance if funding had not yet been secured. Thus, personnel in a lab that receives late notice about the success of its renewal application may alter their career plans even if a break in funding does not ultimately occur.\footnote{Uncertainty is also a function of the score a renewal application received in peer review. That is, the better the score, the more confident a PI will be that their R01 will be funded.} This implies that our estimates, which identify the impacts of funding delays \textit{per se}, are likely smaller than the combined impacts of both uncertainty and funding lapses.

To better understand how PIs perceive and respond to both actual and potential interruptions, we conducted six interviews with PIs who were identified via public grant data as having had an interrupted R01. Every PI we interviewed expressed that funding lapses were a constant worry, even in non-renewal years (one PI simply noted they were ``always'' worried). They stressed the importance of trying to get additional grants (preferably another R01) as a buffer against a potential lapse in funding, so that the salaries of personnel could be shifted to a different grant if necessary. 

They also all spoke about the unpredictability of grant scoring, unpredictability of timing, and lack of communication from the NIH. One PI noted, ``There is a graveyard of grants inside NIH, for every one funded grant, five or six are never funded. Some of my ideas that got funded were sort of lousy, while some very good ideas were not. About 10\% of the grants I write are funded.''

In the event of an interruption, PIs expressed strong aversion to losing personnel as that would be the most disruptive to the functioning of the lab. However, without other grants to compensate, it is difficult to avoid cutting payroll -- the largest grant expense -- in the midst of a funding interruption.\footnote{The only other major expense PIs brought up was animal models (e.g.,~mice).} Non-graduate students are particularly vulnerable to funding interruptions.\footnote{Graduate students are thought to be less vulnerable because universities or departments have commitments to fund their training (e.g.,~they can be shifted to teaching positions even if grant money is unavailable), although this may be less so for graduate students who are in the later stages of their program.} Several PIs expressed regret over having lost ``really good people'' due to interruptions. One PI commented that, ``for me it is like surfing, we have to stay in the front of the wave, and if you get behind it quickly circles down, you don't have people and can't produce data.''

Thinking about the fate of those who leave, one of the PIs noted that ``people who are good get picked up by other labs.'' PIs also mentioned that they tried to time their hiring of people with grant funding cycles, so that postdocs and graduate students would find it natural to leave the lab around the time of a potential funding lapse. Another PI suggested that lab support staff not engaged in the publication process would be first to be let go.

Since interruptions are quite common (about 20\% of the R01s in our sample experience a lapse in funding exceeding 30 days), personnel and research institutions (including the NIH) are well-aware of their possibility and may have developed ways of mitigating their disruptive effects. For example, a PI's home institution may provide bridge funding while a PI waits for delayed funding to arrive. However, the university's willingness to provide support may depend on the belief that external sources will eventually (and preferably quickly) support the lab. While the PIs we interviewed acknowledged that there might be options for bridge funding, these amounts were likely to be small and unlikely to be enough to avoid losing personnel. This accords with remarks from a grants administrator (see Appendix of \citet{tham2023interrupted}) who noted that bridge funding was more likely to be granted for one-time purchases such as equipment rather than ongoing expenses like personnel compensation.

\hypertarget{data}{%
\section{Data}\label{data}}


For our analysis, we need three key pieces of information: (1) which R01 grants were expiring but eventually \emph{successfully renewed}, (2) which personnel were part of labs that depended on those R01s, and (3) the labor market outcomes of those personnel. We obtain these data from: (1) ExPORTER -- a public database of NIH grants, (2) UMETRICS -- administrative grant transaction data from universities (including payments to personnel), and (3) IRS/Census data including the universe of W-2 and 1040 Schedule C (1040-C) tax records and the universe of unemployment insurance (UI) earnings records. Together, these data allow us to identify personnel working in labs with a successfully renewed R01 (but potentially experiencing a funding interruption) and track their entire US employment and earnings history.

\begin{figure}
{\centering \includegraphics{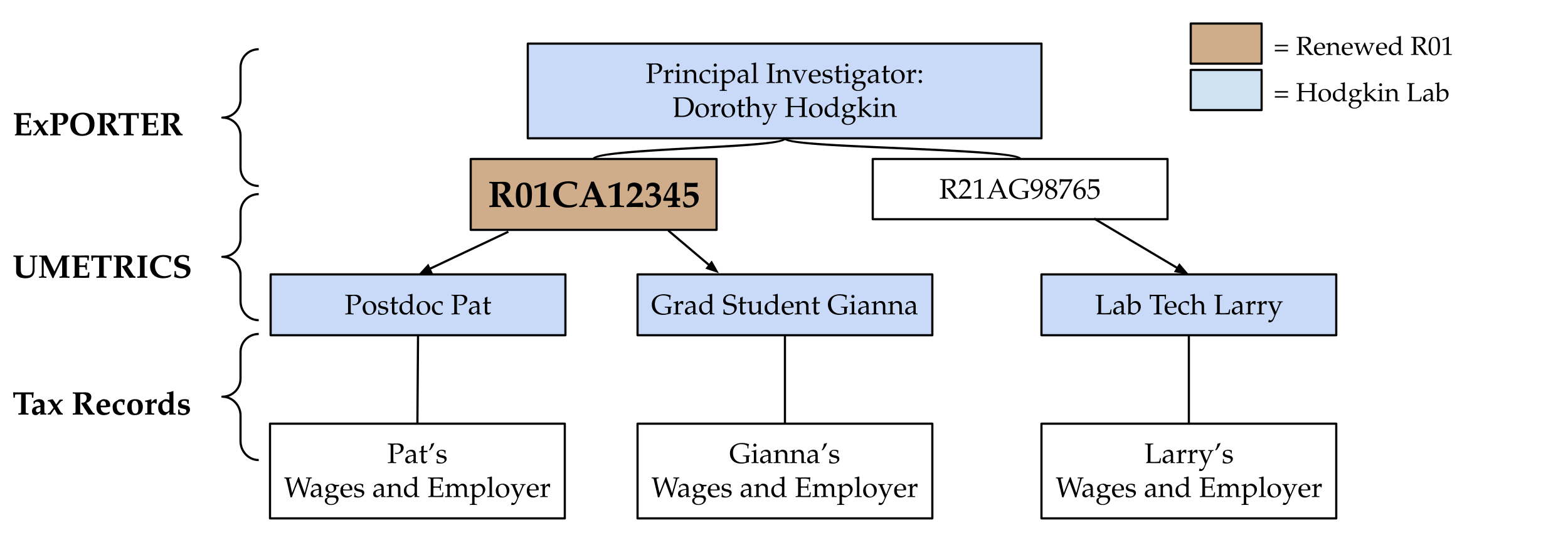} 
}
\caption{\small This diagram shows the process of linking R01 grants to personnel and their labor market outcomes, starting with NIH ExPORTER data at the top and ending with tax and unemployment insurance records stored at the US Census Bureau.}\label{fig:data-construction}
\end{figure}

We illustrate how the data come together with an example. Figure \ref{fig:data-construction} provides a graphical representation.

\begin{enumerate}
\def\labelenumi{\arabic{enumi}.}
\tightlist
\item
  \textbf{Find R01 grants that were successfully renewed.} In this example, we start with an R01 grant with the code R01CA12345 (orange box in Figure \ref{fig:data-construction}). 
\item
  \textbf{Find the PI of the R01.} The PI of R01CA12345 is Dorothy Hodgkin.\footnote{The first British woman to win a Nobel Prize in the sciences.}
\item
  \textbf{Find all of the PI's grants.} Find all grants administered by the PI in the 12 months prior to expiry of the focal R01. In this example, we find that in addition to the focal R01 grant (R01CA12345), PI Hodgkin also had a smaller R21 grant (R21AG98765). 
\item
  \textbf{Find personnel paid by PI's grants.} Find all personnel paid by any of the grants in the 12 months prior to expiry of the focal R01. In this case, PI Hodgkin's two grants were supporting a postdoc, graduate student, and a lab technician. These four personnel (including PI Hodgkin herself) constitute Hodgkins's ``lab''.
\item
  \textbf{Link personnel to labor market outcomes} Merge personnel (including the PI) with IRS/Census earnings and employment data. Link personnel with their employers via the LEHD and W-2 data; employer characteristics come from the LEHD and the LBD; university employers identified using IPEDS data. In this case, the four personnel in the Hodgkins lab are linked to these IRS/Census data sources.
\end{enumerate}


The remainder of the section goes into more detail about these data linkages as well as variable construction. Additional detail is also available in the \protect\hyperlink{data-appendix}{Data Appendix}.

\hypertarget{exporter}{%
\subsection{Grant and PI variables (ExPORTER)}\label{exporter}}

We generate grant- and PI/lab-level variables using ExPORTER, publicly available data on NIH grants provided by the NIH (details in \protect\hyperlink{appendix-exporter}{Data Appendix (ExPORTER)}).\footnote{\url{https://exporter.nih.gov/}}

\textbf{Identifying successfully renewed grants.} We first use ExPORTER to identify \emph{sucessfully renewed} NIH R01 grants -- those that expire and are renewed within the same fiscal year. Since we want to focus on the effects of interruptions, we do not use NIH grants that expire, but are never renewed. 

\begin{figure}
{\centering \includegraphics{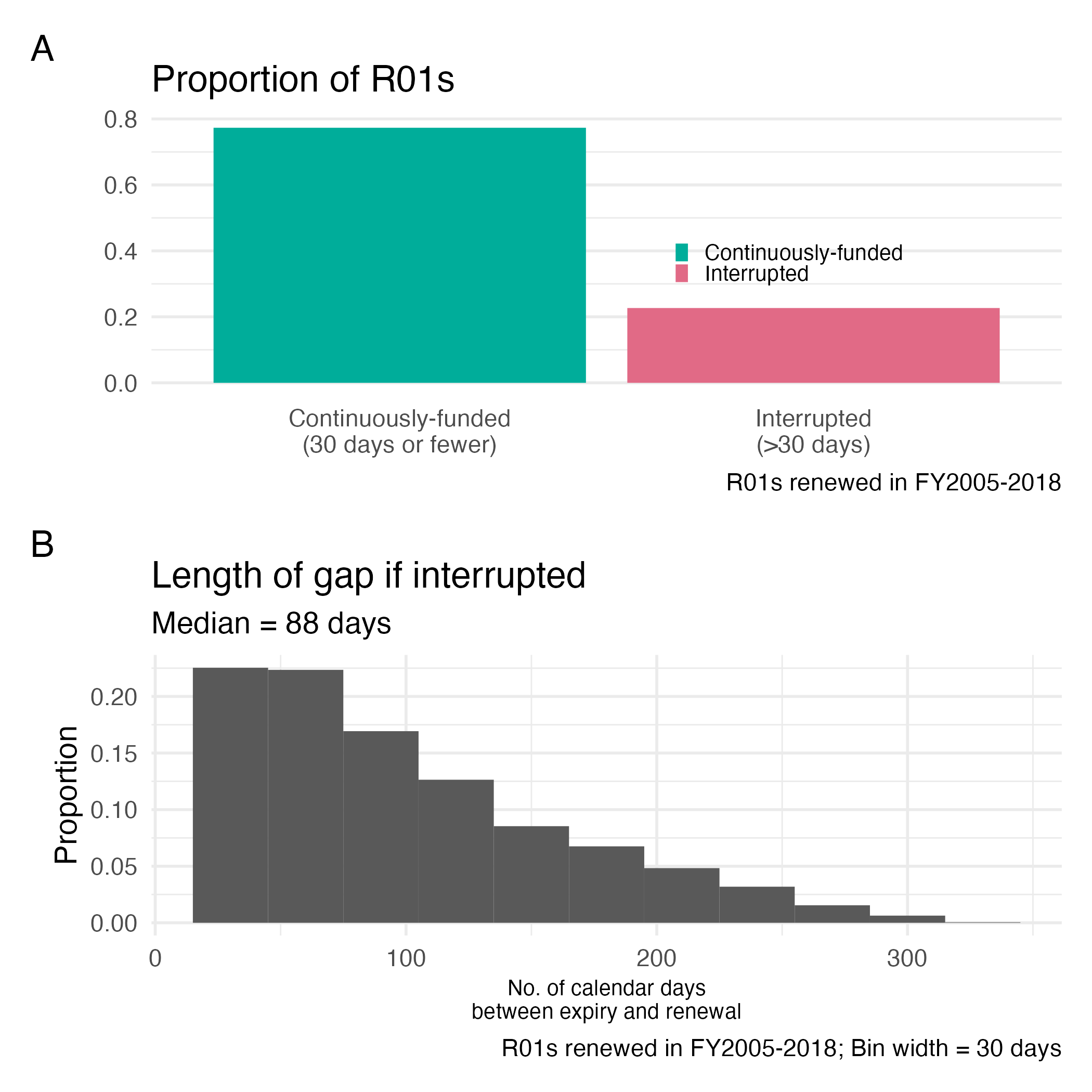} 
}
\caption{\small This figure shows the distribution of funding delays for renewed R01s expiring in Fiscal Years 2005 to 2018. Panel A (top) shows the proportion of R01 grants that are interrupted using our baseline definition of a 30-day delay. Figure B (bottom) shows the distribution of funding delays conditional on being interrupted.}\label{fig:interruption-distribution}
\end{figure}

\textbf{Measuring length of funding gaps and interruptions.} This is measured as the number of calendar days between the end of a project period and the beginning of the next project period. As a baseline, we define an R01 as ``interrupted'' if the gap is 30 or more days (approximating a month as grants are usually disbursed on the first of the month) and as continuously-funded if the gap is fewer than 30 days.\footnote{In Section \ref{sample-split-by-interruption-time}, we examine the effects of interruptions defined by alternative funding gap lengths, finding that longer interruptions (greater than 90 days) do not lead to stronger effects.} Figure \ref{fig:interruption-distribution} shows that, among successfully renewed NIH R01 grants from fiscal years 2005 to 2018, about 20\% were interrupted (Panel A) and the distribution of funding gap lengths conditional on being interrupted is right-skewed with a median of 88 days (Panel B).

\textbf{PI grant portfolio (Number of R01s).} After an interruption, PIs with multiple R01s decrease spending by substantially less than PIs with a single R01 \citep{tham2023interrupted} and PIs we interviewed stressed the fragility of running a lab on only one R01, suggesting that multiple-R01 personnel may be less affected by funding interruptions. Thus, we divide our sample into ``Single-R01'' and ``Multiple-R01'' labs by counting the number of R01s\footnote{More precisely, we define the size of the PI's grant portfolio based on the number of ``R01-equivalent'' grants, including the focal R01. The NIH Glossary states ``R01-equivalent grants are defined as activity codes DP1, DP2, DP5, R01, R37, R56, RF1, RL1, U01 and R35 from select NIGMS and NHGRI program announcements (PAs).'' For brevity, we refer to this variable as the ``Number of R01s'' without explicitly defining the other types of grants included.} a PI has twelve months before and after expiry.\footnote{We include R01s awarded after the focal R01's expiration because these provide a funding cushion if they arrive promptly, and given the time lag between grant application and receiving the funds, the PI would have applied for these grants before the funding status of the focal R01 is known (and so they are not caused by an interruption).} Since post-interruption spending remains relatively stable in multiple-R01 labs, we should see muted effects for the personnel working in these labs.

\hypertarget{umetrics}{%
\subsection{Finding Lab Personnel (UMETRICS)}\label{umetrics}}

We identify personnel who are part of a PI's lab by linking successfully renewed R01 grants from ExPORTER to the UMETRICS database, which contains information from 33 research-intensive universities representing about one-third of US federal research expenditures \citep{umetrics2019review}.\footnote{We use the 2020 release of UMETRICS. All UMETRICS universities are classified as R1 (Doctoral Universities -- very high research activity) according to the Carnegie Classification System and all rank in the top 20\% of universities by federal R\&D expenditures (see Appendix Section \ref{umetrics-herd}).} UMETRICS is administrative transaction-level data on \emph{all} spending from university research grants, including payments from NIH R01 grants to personnel. For each of these payments, we observe the transaction date and the occupation of the personnel at the time of the payment.\footnote{Personnel occupations are assigned by the UMETRICS data team using information such as job titles, and they can change over time (e.g.~a post-doc may become a faculty member).} Thus, we are able to identify the individual research personnel that belong to each PI's lab and their occupation during the run-up to R01 expiry.\footnote{For a given PI with a successfully renewed R01, we identify all NIH grants they administered in the 12 months prior to the focal R01's expiry. We define all personnel paid by any of these grants during the 12 months prior to the focal-R01 expiry as part of the PI's lab.}


\hypertarget{employment-and-earnings-data}{%
\subsection{Employment and Earnings Data}\label{employment-and-earnings-data}}

We use three sources of confidential tax/administrative data, available at the US Census Bureau, to track the earnings of UMETRICS personnel: W-2 tax records, the Longitudinal Employer-Household Dynamics (LEHD) database, and 1040 Schedule C (ILBD) tax records.\footnote{UMETRICS personnel have been linked within Census systems to a confidential person identifier using a probabilistic matching process \citep{wagner2014piks}. This identifier allows us to link UMETRICS personnel to a variety of comprehensive tax, administrative, and survey data held by the US Census Bureau.} We then use confidential information on the universe of US firms from the Longitudinal Business Database (LBD) and the LEHD to identify the characteristics of the employers of these research personnel. Finally, we use public-use information from the Integrated Postsecondary Education Data System (IPEDS) to identify whether the personnel are employed at a university. Appendix Section \ref{appendix-census} provides more details on each of these datasets. With these data, we want to understand how the job placement and earnings of research personnel are affected by funding interruptions. To do so, we construct the variables below.


\textbf{Sector indicators.} We define three mutually exclusive indicators to represent the sectors in which personnel can be employed in a given year:

\begin{enumerate}
\def\labelenumi{\arabic{enumi}.}
\tightlist
\item
  US university (or ``academia'') -- the personnel receives positive earnings from an IPEDS university.
\item
  US non-university (or ``industry'') -- the personnel \emph{only} receives positive earnings from a non-IPEDS US employer.
\item
  Nonemployed in the US (or simply ``nonemployed'') -- the personnel does not receive earnings from an employer in W-2, LEHD, or ILBD data (complement of categories (1) and (2)).
\end{enumerate}


\textbf{Earnings.} We observe yearly earnings for each personnel from 2005 to 2018. These are derived from a combination of W-2, LEHD, and 1040 Schedule C (ILBD) earnings. We define a personnel's total earnings in a given year as their earnings from self-employment (ILBD) plus the maximum of their W-2 and LEHD earnings. That is, \(earnings_{total} = earnings_{ilbd} + max\{earnings_{W2}, earnings_{lehd}\}\).\footnote{The LEHD receives data from individual states unemployment insurance systems and there are two gaps that are particularly important to this study: a) Massachusetts data is not in the LEHD until 2011, and b) graduate student stipends are not covered by unemployment insurance and thus not reflected in LEHD data. The W-2 data fills these gaps.}

\textbf{Presence in the United States (Decennial Census).} To help us distinguish between personnel who are not working but are still present in the US and those who leave the US altogether, we supplement our data with the Decennial Censuses. Each Decennial Census aims to count all people residing in the US on April 1st of the Census year, regardless of nationality, immigration status, or labor force participation. Thus, being observed in the Decennial Census indicates that a personnel was physically present in the US at the time of the Census.\footnote{An exception are Federally Affiliated Count Overseas Operation, who are federal employees (and their dependents) stationed outside the US (mostly military personnel).}




\hypertarget{publication-history}{%
\subsection{Publication History}\label{publication-history}}

Our main outcomes measure the earnings and employment of lab personnel, but these stop at the US border. Publications offer a measure of personnel's scientific activity that is observable no matter where they are employed. We track the publishing activity of UMETRICS personnel by linking them to publications in PubMed, a bibliographic database for biomedical research produced by the US National Library of Medicine (NLM).\footnote{This is done by using an IRIS-provided link between UMETERICS personnel and PubMed publications. The link is built by first disambiguating author names in PubMed and then linking the disambiguated authors to UMETRICS personnel by name, affiliation, ORCID (where available and necessary), email address, and collaborator names.We have also run our analyses using an alternative match to PubMed created by Enrico Berkes and used in \cite{sattari2022ripple}.}

\hypertarget{demographic-data-and-the-decennial-censuses}{%
\subsection{Demographic Data}\label{demographic-data-and-the-decennial-censuses}}

Given immigrants' importance to US science, we explore whether funding delays have heterogeneous effects by personnel place of birth. Place of birth is available along with other demographic information from the Individual Characteristics File (ICF), which is part of the data infrastructure of the LEHD program \citep{vilhuber2014lehd}. Information in the ICF is sourced from the Social Security Administration (SSA) Numident and the Decennial Census.

\hypertarget{analysis-sample}{%
\subsection{Analysis Sample}\label{analysis-sample}}

\begin{table}[htbp]
\centering
\begin{threeparttable}
\caption{Unique Individuals in Single-R01 Labs by Occupation and Birthplace}
\label{tab:sample-n}

\begin{tabular}{lc}
\hline\hline
                      & \textit{Personnel} \\
\textit{Subsample}    & \textit{Count} \\
\hline
All Personnel & 4,200 \\
\addlinespace[0.3em]
\multicolumn{2}{l}{\textbf{\textit{Occupation}}} \\
\hspace{1em}Faculty & 900 \\
\hspace{1em}Postdoc/Grad students & 1,300 \\
\hspace{1em}Others & 2,000 \\
\addlinespace[0.3em]
\multicolumn{2}{l}{\textbf{\textit{Place of Birth}}}\\
\hspace{1em}US-born & 2,700 \\
\hspace{1em}Foreign-born & 1,400 \\
\hline\hline
\end{tabular}

\begin{tablenotes}[flushleft]
\small
\item \hspace{-0.25em}This table shows the breakdown, by occupation and place of birth, of personnel belonging to single-R01 labs. Due to rounding required by Census disclosure avoidance rules, the summation across categories may not always equal the total.
\end{tablenotes}
\end{threeparttable}
\end{table}

\begin{table}[htbp]
\begin{threeparttable}
\caption{\label{tab:balance-tab} Covariate balance between interrupted and continuously-funded labs}

\begin{tabular}{>{\raggedleft\arraybackslash}p{4cm}>{\raggedleft\arraybackslash}p{2.5cm}>{\raggedleft\arraybackslash}p{2.5cm}>{\raggedleft\arraybackslash}p{2.5cm}>{\raggedleft\arraybackslash}p{2.5cm}}
\hline\hline
\vspace{-12pt} \\
Variable & Difference in means & Continuously Funded & Interrupted & P-value\\
\midrule
\addlinespace[0.3em]
\multicolumn{5}{l}{\textbf{Panel A: Single-R01 Labs}}\\
\addlinespace[0.3em]
\multicolumn{5}{l}{\textit{Lab size \& composition}}\\
\hspace{1em}\hspace{1em}\% Faculty & -0.02 (0.02) & 0.33 & 0.31 & 0.36\\
\hspace{1em}\hspace{1em}\% Postdoc/Grad Student & 0.01 (0.03) & 0.28 & 0.29 & 0.82\\
\hspace{1em}\hspace{1em}\% Other Occ & 0.01 (0.03) & 0.39 & 0.40 & 0.60\\
\hspace{1em}\hspace{1em}Lab Size & 0.75 (1.18) & 7.72 & 8.47 & 0.52\\
\hspace{1em}\hspace{1em}\% Female & -0.02 & 0.46 & 0.44 & \\
\hspace{1em}\hspace{1em}\% Asian & 0.02 & 0.27 & 0.29 & \\
\hspace{1em}\hspace{1em}\% Black & 0.01 & 0.02 & 0.03 & \\
\hspace{1em}\hspace{1em}\% White & -0.03 & 0.67 & 0.65 & \\
\hspace{1em}\hspace{1em}\% Hispanic & -0.01 & 0.05 & 0.04 & \\
\hspace{1em}\hspace{1em}\% US-Born & 0.01 & 0.56 & 0.57 & \\
\addlinespace[0.3em]
\multicolumn{5}{l}{\textit{Pubs \& Funding}}\\
\hspace{1em}\hspace{1em}Pubs per year & 3.38 (1.74) & 9.05 & 12.43 & 0.05\\
\hspace{1em}\hspace{1em}NIH Funding (millions per year) & -0.02 (0.04) & 0.50 & 0.48 & 0.65\\
\addlinespace[0.3em]
\multicolumn{1}{l}{\textit{Personnel Count}} &  & 667 & 182 & \\
\addlinespace[0.3em]
\hline
\addlinespace[0.3em]
\multicolumn{5}{l}{\textbf{Panel B: Multiple-R01 Labs}}\\
\addlinespace[0.3em]
\multicolumn{5}{l}{\textit{Lab size \& composition}}\\
\hspace{1em}\hspace{1em}\% Faculty & 0.01 (0.01) & 0.30 & 0.30 & 0.61\\
\hspace{1em}\hspace{1em}\% Postdoc/Grad Student & 0.00 (0.01) & 0.29 & 0.29 & 0.96\\
\hspace{1em}\hspace{1em}\% Other Occ & -0.01 (0.01) & 0.41 & 0.41 & 0.68\\
\hspace{1em}\hspace{1em}Lab Size & 1.99 (1.56) & 14.96 & 16.95 & 0.20\\
\addlinespace[0.3em]
\multicolumn{5}{l}{\textit{Pubs \& Funding}}\\
\hspace{1em}\hspace{1em}Pubs per year & 0.59 (1.74) & 19.16 & 19.75 & 0.74\\
\hspace{1em}\hspace{1em}NIH Funding (millions per year) & -0.07 (0.06) & 1.18 & 1.11 & 0.24\\
\addlinespace[0.3em]
\multicolumn{1}{l}{\textit{Personnel Count}} &  & 1,313 & 417 & \\
\hline\hline
\end{tabular}

\begin{tablenotes}[flushleft]
\small
\item \hspace{-0.25em}This table shows differences in means, across lab characteristics, for interrupted and continuously-funded labs at the time of R01 renewal, by whether the labs have a single R01 (Panel A) or multiple R01s (Panel B). The unit of observation is a PI/lab-by-R01 renewal. Publications and funding are average publications and funding per year for the five years prior to R01 renewal. Due to Census disclosure avoidance restrictions, we only show the means of demographic characteristics for single-R01 labs. Standard errors and p-values are from a two-sided t-test. 
\end{tablenotes}
\end{threeparttable}
\end{table}

Our final sample consists of about 4,200 research personnel belonging to 600 single-R01 labs and about 13,500 personnel belonging to 1,200 multiple-R01 labs. Using information from UMETRICS (Section \ref{umetrics}), we define three occupations: faculty, trainees (postdocs and graduate students), and ``others'' (which includes occupations such as staff, research scientists, and undergrads). Using information from the ICF (Section \ref{demographic-data-and-the-decennial-censuses}), we classify personnel by whether they are born in the US (``US-born'') or not (``foreign-born'').\footnote{Although ``non-US-born'' is more accurate, we use the term ``foreign-born'' for easier reading.} Table \ref{tab:sample-n} shows that about 21\% of lab personnel are faculty, 31\% are postdocs or graduate students, and the remaining 48\% have other occupations. About two-thirds of lab personnel were born in the US, with the remaining third born elsewhere.


Table \ref{tab:balance-tab} compares interrupted and continuously-funded labs across a range of characteristics, including lab size and composition, number of publications five years prior to R01 renewal, and amount of funding five years prior to R01 renewal. There are no statistically significant differences, and the only substantive difference is that interrupted single-R01 labs produce about three more publications per year than their continuously-funded counterparts. However, to the extent that publications measure lab ``quality'', this difference indicates that lower quality labs do not select into being interrupted. 
Though not necessary for the parallel trends assumption to hold, the similarity of these pre-expiry baseline characteristics shows that interrupted and continuously-funded personnel are working in otherwise comparable labs and increases the plausibility that they would have been on parallel paths if not for the interruption.

\hypertarget{estimation}{%
\section{Estimation}\label{estimation}}

\hypertarget{stacked-difference-in-differences}{%
\subsection{Stacked Difference-in-Differences}\label{stacked-difference-in-differences}}

We use a difference-in-differences (DiD) design to estimate the effect of funding interruptions on personnel. Our strategy involves three steps:

\begin{enumerate}
\def\labelenumi{\arabic{enumi}.}
\tightlist
\item Stack the data by event-year (i.e.,~the focal R01's expiration year).
\item Identify continuously-funded personnel within each expiration year cohort that have not experienced an interruption in a two-year window around that expiry year (i.e.,~identify ``clean controls'').
\item
  Estimate average treatment effects (ATTs) and event studies using a modified version of the \cite{callaway2020dd} (CS) estimator that compares interrupted and ``clean'' continuously-funded personnel \emph{within the same R01 expiry year}.
\end{enumerate}


In a typical staggered DiD setting, the CS estimator estimates disaggregated ``group-time'' treatment effects, where groups are defined by time of treatment.\footnote{These group-time effects can then be aggregated as desired (e.g.,~as a static treatment effect or by time relative to treatment for an event study).} In that setting, control units are not assigned to groups because they do not have a defined counterfactual treatment period.\footnote{In many applications it is difficult to define a counterfactual treatment time. For instance, the year in which a state might have but did not pass a minimum wage increase.} In contrast, in our setting, since every R01 has a clear expiration year, all personnel (whether interrupted or continuously-funded) have a well-defined treatment period. Thus, in Step (1), take advantage of this data structure by stacking personnel by R01 expiry year cohorts, forming a cohort-by-personnel-by-time panel dataset. 
Each cohort can be thought of as a separate DiD/event study with a single treatment period. With this structure, we are more likely to compare personnel in labs with projects and budgets that are at similar stages in their lifecycle.\footnote{For instance, personnel may be less likely to leave their job at the beginning of an R01 than at the end of an R01, so using them as control units will overstate the effect of an interruption.}

After stacking, we limit control units (continuously-funded personnel) within a given expiry-year cohort to those that are ``clean'' -- that is, they are not treated in a time window of interest. Specifically, in Step (2) we require that, to be included in an expiry-year cohort, control personnel must not be treated two years before or after the expiration year of the focal R01.\footnote{For instance, in the cohort with expiration year 2001, control (continuously-funded) personnel must not have been treated in any year from 1999 to 2003. This would not be the case if personnel are in a lab with two R01s that are expiring in consecutive years, 2001 and 2002. The first R01 is continuously-funded in 2001, but the second R01 is interrupted in 2002. Without any restrictions, these personnel would be controls in the 2001 cohort but treated in 2002.}

With our stacked data structure, it is common to use a modified two-way fixed effects estimator with unit-cohort fixed effects and time-cohort fixed effects \citep{baker2022dd}. In Step (3), we instead use the CS estimator because it allows for transparent and flexible aggregation of the group-time treatment estimates.\footnote{CS aggregates treatment effects by group size. Two-way fixed effects implicitly uses OLS weights which is more efficient at the cost of bias \citep{baker2022dd}.}



As discussed in Section \ref{exporter}, evidence on lab spending\footnote{We show in \cite{tham2023interrupted} that funding interruptions substantially reduce grant expenditures in single-R01 labs but only modestly reduces spending (in log points) in multiple-R01 labs (see Appendix Figure \ref{fig:lab_spending}).} and PI interviews suggest that personnel can easily be moved between a lab's grants. Thus, personnel working in labs supported by multiple R01s may be cushioned from the effects of a funding interruption. This motivates us to estimate the effects of funding interruptions on career outcomes separately for personnel in multiple- and single-R01 labs. 


\hypertarget{benchmark-main}{%
\subsection{Benchmarking Estimates}\label{benchmark-main}}

As discussed in Section \ref{employment-and-earnings-data}, many of our outcome variables are indicators (e.g.,~sector of employment), and so our estimates reveal percentage point (pp) changes after an interruption. To better understand the magnitude of these estimates, we benchmark them  against the absolute change in the mean of the corresponding outcome for the control group (i.e.,~personnel in continuously-funded labs), just before treatment ($\bar{y}^{c}_{-1}$) and five years post-treatment ($\bar{y}^{c}_{5}$). That is,
\begin{equation*}
    y_{benchmark} = |\bar{y}^{c}_{5} - \bar{y}^{c}_{-1}|.
\end{equation*}
This change gives us a sense of how an outcome would have evolved in the absence of an interruption. We then report the size of an estimate (either an event study coefficient or an aggregated average treatment effect (ATT)), denoted $\hat{\beta}$, as a percentage of $y_{benchmark}$. That is,
\begin{equation*}
   \text{Effect size} = \frac{\hat{\beta}}{y_{benchmark}}.
\end{equation*}

This reveals the size of an interruption impact on an outcome relative to the underlying change of that outcome.

\hypertarget{identification}{%
\subsection{Identification}\label{identification}}

We rely on two main assumptions to identify the average treatment effect on the treated (ATT) of an interruption on the career outcomes of personnel: (1) parallel trends and (2) no anticipation.

\textbf{Parallel Trends.} The parallel trends assumption requires that the average outcome among the treated and comparison populations would have followed parallel trends in the absence of treatment. In our context, this means that the employment and earnings outcomes for interrupted and continuously-funded personnel would have evolved in parallel if the funding interruption had not occurred.

The parallel trends assumption allows treatment to be non-random based on characteristics that affect the \emph{level} of the outcome but requires the treatment be mean independent of characteristics that affect the \emph{trend} of the outcome. For instance, highly organized PIs may select into the continuously-funded control group because they are more likely to submit their grant renewal paperwork on time and avoid a funding interruption. Their high level of organization may also affect employee outcomes (e.g.,~by ensuring that postdocs and graduate students are regularly publishing in a timely manner). However, as long as unobservable PI organizational skills affect employee outcomes in the same way both before and after the treatment, it does not violate the parallel trends assumption.

\begin{figure}
{\centering \includegraphics[width=0.93\linewidth]{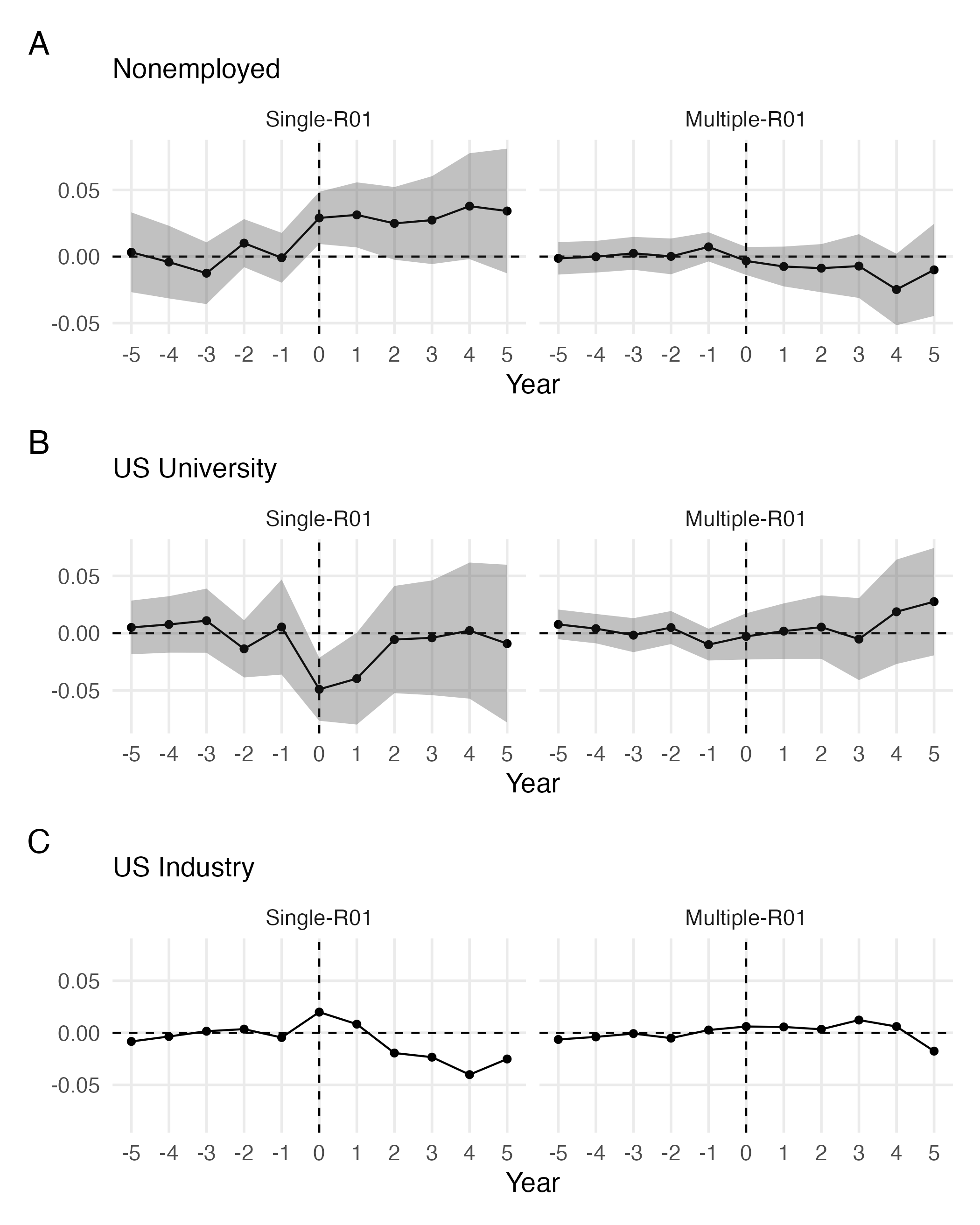} 
}
\caption{\small This figure shows event studies of the effects of interruptions on three mutually exclusive employment outcomes: 1) nonemployed in the US (Panel A), 2) employed at a US university (Panel B), or 3) employed in US industry (Panel C). They are obtained using our modified \cite{callaway2020dd} estimator. The expiration of a lab's grant takes place at year 0, and the estimated interruption effects range from 5 years before to 5 years after expiry. The left column is for personnel in a single-R01 lab and the right column is for personnel in a multiple-R01 lab. Standard errors are bootstrapped and clustered at the expiring-R01-level. Standard errors not available for Figure C for Census disclosure avoidance reasons. The data underlying this graph can be found in Table \ref{tab:eventstudies_sector}.}\label{fig:es-sector}
\end{figure}
\begin{figure}
{\centering \includegraphics[width=0.95\linewidth]{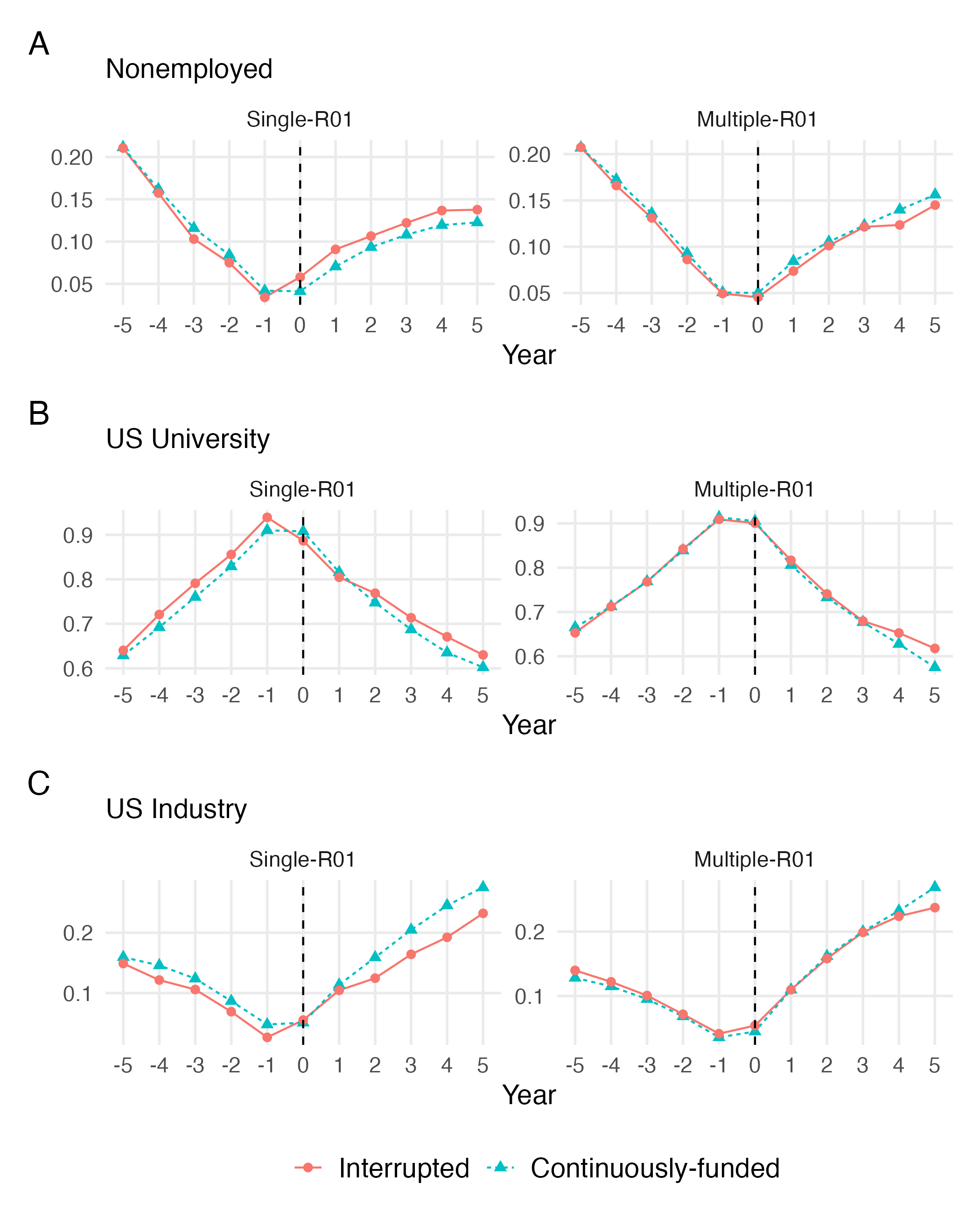}}
\caption{\small This figure shows the average probability that a personnel is in one of three mutually exclusive employment categories: 1) nonemployed in the US (Panel A), 2) employed at a US university (Panel B), or 3) employed in US industry (Panel C). These probabilities are calculated from five years before R01 grant expiration to five years after. Figures in the left column are for personnel in a single-R01 lab and figures in the right column are for personnel in a multiple-R01 lab. Both groups are split into continuously-funded and interrupted labs. The data underlying this graph can be found in Table \ref{tab:rawmeans}.}\label{fig:fraction-by-sector}
\end{figure}
\begin{figure}
{\centering \includegraphics[width=0.9\linewidth]{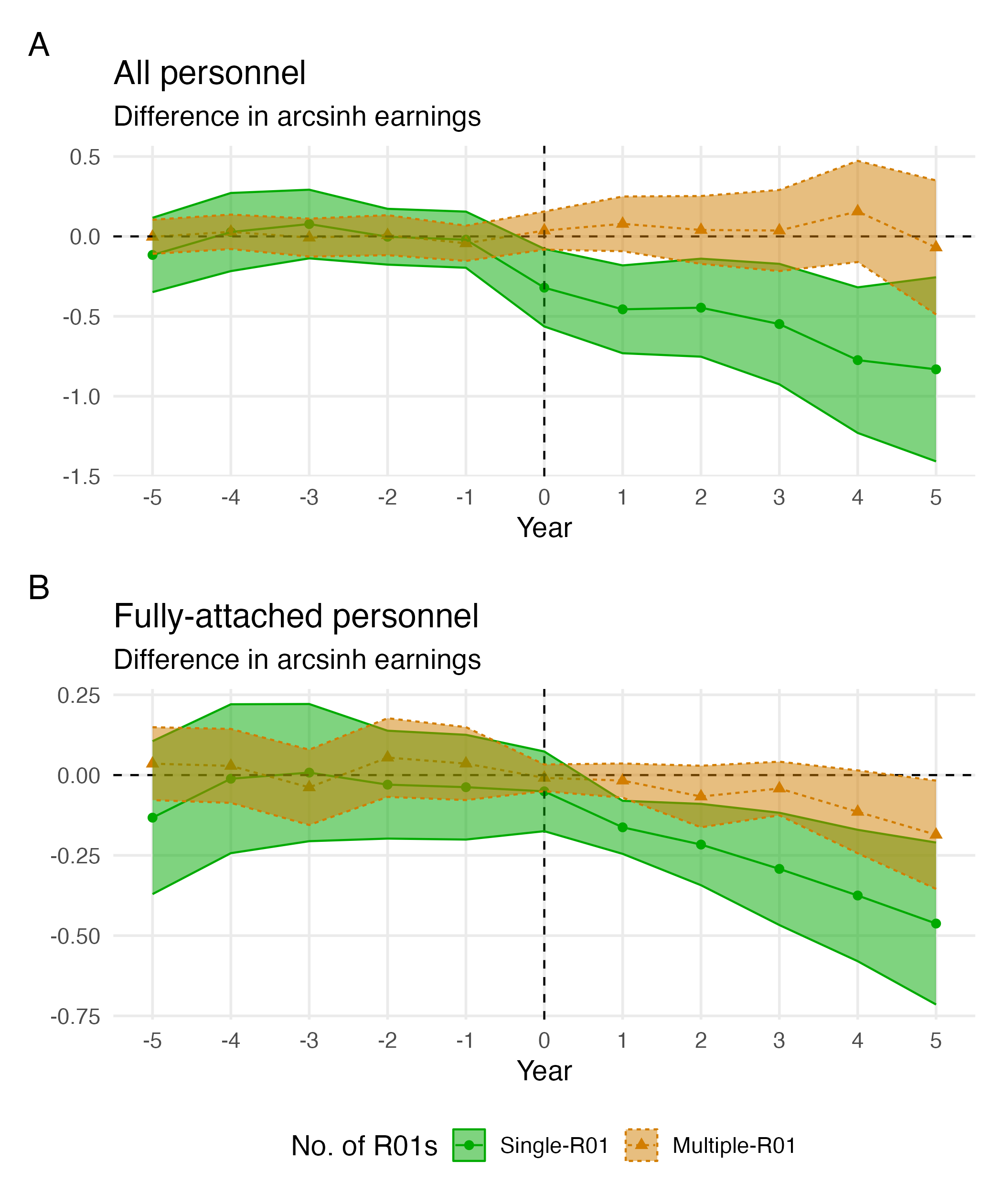} 
}
\caption{\small This figure shows event studies of the effects interruptions on the arcsinh of total earnings. They are obtained using our modified \cite{callaway2020dd} estimator. The expiration of a lab's grant takes place at time 0, and the estimated interruption effects range from 5 years before to 5 years after expiry. The green series is for personnel in a single-R01 lab and the orange series is for personnel in a multiple-R01 lab. The top graph (Panel A) is for the full sample of all research personnel and the bottom graph (Panel B) is for the subsample of research personnel that have positive earnings in all periods from one period before expiry to five years after expiry (i.e., the fully attached subsample). Standard errors are bootstrapped and clustered at the expiring-R01-level. The data underlying this graph can be found in Table \ref{tab:eventstudies_earnings}.}\label{fig:es-wages}
\end{figure}

Though the parallel trends assumption cannot be tested, there are several reasons it is plausible in the setting of R01 interruptions. First, neither raw means (Figure \ref{fig:fraction-by-sector}) nor event studies (Figures \ref{fig:es-sector} and \ref{fig:es-wages}) show evidence of diverging trends prior to grant expiration. Second, balance statistics (Table \ref{tab:balance-tab}) suggest that interrupted and continuously-funded labs are quite similar across a variety of characteristics, including demographics (gender, race, ethnicity, and place of birth), occupational composition, and research production. Third, the effects of interruptions are concentrated among non-faculty in labs supported by a single R01, which is who, \textit{ex ante,} we would expect to bear the brunt of funding delays. Moreover, if time-varying unobserved confounders were driving our results, they presumably affect multiple-R01 labs in ways that are similar to how they affect single-R01 labs. However, in contrast to single-R01 labs, we find little evidence that the employment outcomes of personnel in multiple-R01 labs are impacted by interruptions.

\textbf{No or Limited Anticipation.} The no anticipation assumption requires zero treatment effect prior to the treatment actually taking place. 
The limited anticipation assumption relaxes this requirement if anticipation occurs at a fixed length of time before the treatment. In this case, the treatment period can be redefined as the point when units are aware of treatment. We do not see any differences between treated and control outcomes before treatment, suggesting that personnel and their PIs do not differentially anticipate interruptions more than a year in advance (our data are at a yearly frequency). 

\hypertarget{results}{%
\section{Results}\label{results}}

\hypertarget{employment-outcomes}{%
\subsection{Sector placement}\label{employment-outcomes}}

In this section, we consider whether funding delays affect personnel's sector of employment. We construct three mutually exclusive outcomes based on the observed Employer Identification Numbers (EIN) in personnel's W-2s and the LEHD. In a given year, an individual can be (1) not employed in the US (``nonemployed''), (2) employed at a US university (``academia''), (3) employed in the US but not at a university (``industry'').

\textbf{Event Studies.} Figure \ref{fig:es-sector} shows event studies for these three employment outcomes, estimated using the stacked difference-in-differences method described in Section \ref{stacked-difference-in-differences}. The estimates are plotted for years -5 to 5, where year 0 is the year of R01 grant expiry. 

The left column of graphs show estimates for personnel in single-R01 labs. After an interruption, there is an immediate and persistent 3~pp increase in nonemployment. Personnel also shift sectors within the US. There is an immediate 5~pp decrease in the probability of being at a US university, but after two years continuously-funded personnel ``catch up'' in their university departures (as seen in the raw means below), so the two groups are equally likely to be working at a US university. Correspondingly, interrupted personnel are initially 2~pp more likely to be in US industry, but become 2~pp less likely after two years.

Overall, these patterns suggest the following. First, the level shift in nonemployment indicates a one-time move out of the US labor force for interrupted personnel. Second, the initial decrease and then recovery in US university employment suggests a shift in the timing of departures -- that is, interruptions induced personnel to leave the university sector earlier than they otherwise would have -- but did not change the long-term stock of personnel at US universities.

In contrast, for personnel in multiple-R01 labs (right column of graphs), interruptions have no effect on any of the three employment outcomes. This suggests that multiple funding sources insulate lab personnel from the consequences of an interruption.\footnote{See Appendix Section \ref{alternative-control-group} for a direct comparison of the two interrupted groups.}

\textbf{Raw Means.} To unpack the mechanics underlying our event studies, Figure \ref{fig:fraction-by-sector} displays the raw fractions of personnel that fall into each employment category (separately for interrupted and continuously-funded labs). The left and right columns of graphs display means for single- and multiple-R01 personnel, respectively.

We first see that, even unconditionally, the mean employment outcomes of interrupted and continuously-funded personnel trend similarly prior to R01 expiry, which is again consistent with a parallel trends assumption. Second, the post-expiry relative changes in these unconditional means closely mirror the changes observed in the event studies of Figure \ref{fig:es-sector}; these raw means allow us to see the separate dynamics of both interrupted and continuously-funded personnel.\footnote{For single-R01 personnel (left column of graphs), the fraction of nonemployed interrupted personnel increases immediately (in year 0) and persists until at least year 5. At the same time, a temporary initial increase in interrupted personnel's rate of exit from universities is later offset by a relatively slower exit rate, allowing the interrupted to return to baseline. Inversely, the interrupted initially enter industry more quickly, but a relative slowdown later returns them to baseline. For multiple-R01 personnel (right column of graphs), the employment outcomes of the interrupted and continuously-funded are similar (in both levels and trends) before and after R01 expiry.} Finally, we observe that our three outcomes are either U- or inverted U-shaped, which is largely driven by our sample requirement that research personnel receive payments from a grant at their UMETRICS university during the 12 months prior to R01 expiry (see Section \ref{exporter}).\footnote{The probability of being nonemployed in the US (Figure \ref{fig:fraction-by-sector}A) starts at 20\%, declines to 5\% around grant expiry, and then increases to 15\% five years later. The probability of being at a US university Figure (\ref{fig:fraction-by-sector}B) starts at 65\% in year -5, peaks at 90\% in year 0, and declines to 60\% in year 5. The probability of US industry employment (Figure \ref{fig:fraction-by-sector}C) starts at 15\%, declines to 5\% around grant expiry, then increases to over 20\% five years later. See Table \ref{tab:rawmeans}.} In particular, the inverted U-shape for university employment makes it clear that the recovery visible in the left column event study of Figure \ref{fig:fraction-by-sector}B occurs because interrupted personnel leave universities \emph{earlier} than the continuously-funded -- however, both groups flow out of universities after grant expiry.

Overall, the raw means, like the event studies, clearly suggest that funding interruptions to single-R01 labs substantially alter employment patterns for the personnel of those labs, affecting entrance to nonemployment and the \emph{timing} of departures from universities to industry. 

\textbf{Aggregated Static Effects.} Table \ref{tab:tab-main-subsample} presents the post-treatment estimates for the three employment outcomes, aggregated over the entire five year post-expiry period. Single-R01 personnel experience a 3~pp (37\%)\footnote{As noted in Section \ref{benchmark-main}, we benchmark our percentage point effects against the absolute change in the mean of the corresponding outcome for continuously-funded personnel, just before treatment and five years post-treatment. In this case, 4.2\% of continuously-funded personnel are nonemployed one year prior to grant expiry and this increased to 12.3\% five years after expiry. Thus, we compute the effect size as 0.03/|0.123 - 0.042| = 0.37. We perform a similar calculation for other outcomes.} increase in the probability of nonemployment, which confirms the seemingly permanent exit of these individuals from the US labor market. The catch-up visible in the event studies of Figure \ref{fig:es-sector} (and in the raw means of Figure \ref{fig:fraction-by-sector}) leads to a statistically insignificant drop in the longer-term probability of employment in US academia, despite the sharp and statistically significant initial drop. We again observe that the employment outcomes of multiple-R01 personnel are unaffected by interruptions.

\begin{table}[htbp]
\centering
\begin{threeparttable}
\caption{\label{tab:tab-main-subsample}Effect of Interruptions on Main Employment Outcomes}

\begin{tabular}{lccc} \hline\hline
& Nonemployed & US University & US Industry  \\
\textit{Subample} & (1) & (2) & (3) \\
\hline
\vspace{-12pt} \\
\textbf{Panel A: Single-R01 Labs} \\
\\
All Personnel                & 0.0302*** & -0.0210 & -0.0092 \\
                             & (0.0117) & (0.0184) & (0.0147) \\
\\
\textbf{\textit{Place of Birth}} \\
\hspace{3mm}Foreign-Born                & 0.0183 & 0.0109 & -0.0292* \\
                                        & (0.0231) & (0.0267) & (0.0158) \\
\vspace{-12pt} \\
\hspace{3mm}US-Born                      & 0.0350*** & -0.0354* & 0.0007 \\
                                         & (0.0111) & (0.0210) & (0.0195) \\
\\
\textbf{\textit{Occupation}} \\
\hspace{3mm}Faculty                      & 0.0030 & -0.0213 & 0.0216 \\
                                         & (0.0152) & (0.0178) & (0.0150) \\
\vspace{-12pt} \\
\hspace{3mm}Postdoc/Grad                 & 0.0610*** & -0.0537** & -0.0073 \\
                                         & (0.0235) & (0.0310) & (0.0240) \\
\vspace{-12pt} \\
\hspace{3mm}Other                        & 0.0262* & -0.0075 & -0.0187 \\
                                         & (0.0145) & (0.0287) & (0.0246) \\
\\
\hline
\vspace{-12pt} \\
\textbf{Panel B: Multiple-R01 Labs} \\
\\
All Personnel                & -0.0093 & 0.0056 & 0.0037 \\
                             & (0.0086) & (0.0133) & (0.0101) \\
\vspace{-12pt} \\
\hline\hline
\end{tabular}









\begin{tablenotes}[flushleft]
\small
\item \hspace{-0.25em}This table shows aggregated estimates of the average treatment effects (ATTs) of an interruption on three mutually exclusive employment outcomes: 1) nonemployed in the US, 2) employed at a US university, or 3) employed in US industry. They are obtained using our modified \cite{callaway2020dd} estimator. The effects are aggregated over the 5 years after the expiration of a lab's grant. Panel A shows the results for personnel in a single-R01 lab and Panel B shows the results for personnel in multiple-R01 labs. Standard errors are bootstrapped and clustered at the expiring-R01-renewal level. *, **, and *** indicate statistical significance at the 10\%, 5\%, and 1\% levels. 
\end{tablenotes}
\end{threeparttable}
\end{table}


\textbf{Heterogeneity by Birthplace.} The scientific workforce in the US is heavily dependent on immigration. In our sample, slightly more than a third of lab personnel were born outside the US (compared to 13\% for the overall US population),\footnote{As estimated by the American Community Survey, see  \href{https://data.census.gov/table/ACSDP5Y2018.DP02}{ACS Table DP02, 2018 5-year estimates} and \href{https://data.census.gov/table/ACSDP5Y2010.DP02}{ACS Table DP02, 2010 5-year estimates}.} and visa restrictions may constrain their job choices. To better understand how these restrictions interact with funding interruptions, we split the sample into US-born personnel who face no visa-related work restrictions and personnel born outside the US (``foreign-born'') who are likely to be on a visa.\footnote{Some of the foreign-born could be naturalized citizens or permanent residents; we cannot observe this in our data.}

Table \ref{tab:tab-main-subsample} shows that the main employment effects are primarily concentrated among the US-born. Post interruption, the single-R01 US-born are 3.5~pp (60\%) more likely to be nonemployed in the US. In contrast, the foreign-born are only 1.8~pp (13.5\%; statistically insignificant) more likely to be nonemployed in the US. The large effect for the US-born is almost entirely driven by departures from universities -- there is a symmetric 3.5~pp (11.7\%)  decrease in the probability of working at a US university and no change (0.1~pp) in the probability of working in US industry. Thus, it is primarily the US-born that are induced by funding interruptions to depart from US universities. 




It is perhaps surprising that we find larger nonemployment effects for the US-born.\footnote{It is important to emphasize that the unconditional probability of being nonemployed in the US is much higher for the foreign born (19\% in year 5) than the US-born (9\% in year 5).}Foreign-born personnel on visas face significant restrictions on both their ability to remain in the US without a job, and on their ability to gain access to US employment from abroad, which in some cases may lead them to have stronger preferences for a US job.\footnote{For example, \cite{ganguli2019postdocs} find a stronger preference among foreign-born graduate students in US PhD programs for a US-based postdoc position.} At the same time, universities have access to visa categories unavailable to private employers, making it easier for foreign-born personnel to remain in the US by working at a university. Consistent with this, the point estimates in Table \ref{tab:tab-main-subsample} for the foreign-born are positive for the university employment outcome and negative for the industry employment outcome. In contrast, the point estimates are reversed for the US-born.\footnote{Appendix Table \ref{tab:tab-censusonly} also hints at visa restrictions on movement, showing the impacts of interruption on the probability of receiving positive university earnings from a personnel's own UMETRICS university or another university.  The point estimate on US-born personnel working at their own university is negative ($-3.5$~pp), suggesting they are pushed to leave their UMETRICS university. In contrast, the point estimate for interrupted foreign-born personnel is positive (1.8~pp), suggesting that interruptions may lead to less job mobility among foreign-born personnel.} In addition, though the US-born are more likely to be driven to nonemployment, \emph{conditional} on being induced into nonemployment, the foreign-born are more likely to stay outside the US (see Section \ref{presence-in-the-united-states} below).

Thus, the US-born again appear notably more mobile than their foreign-born counterparts, able to more easily leave their current employment situation, moving universities, sectors, or even become nonemployed in the US. Though the choice of foreign-born to stay in academia rather than become nonemployed or enter the non-university sector is not necessarily due to constraints imposed by visa and work authorizations, overall our results suggest foreign-born personnel face significant job mobility constraints.

\textbf{Heterogeneity by Occupation.} Contractual arrangements vary dramatically across different types of personnel within a lab. Typically, the faculty PI is on a permanent contract with the university\footnote{There is variation even among faculty contracts. In the biomedical sciences, faculty are often on ``soft money'' contracts, where some part or even all of their salary is funded by external grants, with no guarantee of salary from the university, even in exchange for teaching. Other faculty, particularly in medical schools, might be asked to raise their salaries either through grants or through clinical work, that is, seeing patients in an associated hospital.} and is supported by university money, while postdocs are usually on temporary contracts (with no renewal guarantee) and are supported by grant funding. Graduate students have a finite horizon at the university, but typically enjoy employment guarantees (in a teaching, if not a research position), though these commitments likely vary across universities and may weaken as graduation approaches. Staff contracts also vary, with some supported by grants and on year-to-year contracts, and others on permanent university contracts. These contractual differences suggest that funding interruptions may disproportionately impact occupations with a less permanent relationship with the university.


Table \ref{tab:tab-main-subsample} shows that the main employment effects for single-R01 labs are primarily driven by trainees (graduate students and postdocs). After an interruption, trainees in single-R01 labs are 6.1~pp (60\%) more likely to be nonemployed in the US. As with the US-born, this large effect is almost entirely driven by departures of these trainees from universities. Indeed, there is a 5.4~pp (15\%) decrease in probability of working at a university and no change (-0.7~pp) in the probability of working in industry. In contrast, employment effects for faculty and other occupations are muted, consistent with many of them being on permanent contracts.

\hypertarget{presence-in-the-united-states}{%
\subsection{Presence in the United States}\label{presence-in-the-united-states}}

The post-interruption increase in the probability of US nonemployment for single-R01 personnel raises the question of what these individuals are doing, since we have so far only been able to say what they are \emph{not} doing -- namely working for pay in the United States. These individuals may still be physically present in the US, or they may have left the country entirely. Since the scientific workforce is relatively mobile, leaving the US to find a research job in another country is particularly plausible. 

To address this, we use Decennial Censuses, which strive to enumerate every person living in the United States and thus allow us to observe the actual presence of personnel in the US, regardless of their employment status. Specifically, we create two new outcome variables. First, an indicator taking a value of one if a person is nonemployed in the US \emph{and} is \emph{absent} from the 2020 Decennial Census (zero otherwise). Second, an indicator taking a value of one if a person is nonemployed in the US \emph{and} is \emph{present} in the 2020 Decennial Census (zero otherwise). 

Table \ref{tab:tab-noadmin-census-subsample-maintext} presents our estimates for single-R01 labs. 
After an interruption, personnel are 1.6~pp more likely to be nonemployed and absent from the 2020 Decennial Census and are 1.5~pp more likely to be nonemployed but present in the 2020 Decennial Census. These estimates imply that slightly more than half ($1.6/(1.5+1.6)$) of the personnel induced to nonemployment by an interruption leave the US permanently, while the rest either leave for a time and return by mid-2020 or have been present in the US but not working. Thus, it appears that a substantial fraction of personnel displaced by an interruption end up leaving the US altogether.\footnote{In Appendix Table \ref{tab:tab-censusonly}, we use three consecutive Censuses -- 2000, 2010, and 2020 -- to estimate the impact of an interruption on the probability of personnel being present in each Census. We find that single-R01 personnel are about 3~pp less likely to be observed in a Decennial Census after an interruption.}



\begin{table}[htbp]
\centering
\begin{threeparttable}
\caption{\label{tab:tab-noadmin-census-subsample-maintext}Effect of Interruptions on Presence in the 2020 Decennial Census for Nonemployed Personnel from Single-R01 Labs}

\begin{tabular}{lcc} \hline\hline
& Nonemployed & Nonemployed   \\
& In 2020 Census & Not In 2020 Census   \\
\textit{Subample} & (1) & (2)  \\
\hline
\vspace{-12pt} \\
All Personnel                           & 0.0146 & 0.0156** \\
                                        & (0.0096) & (0.0067) \\
\\
\textbf{\textit{Place of Birth}} \\
\hspace{3mm}Foreign-Born                & -0.00887 & 0.0282 \\
                                        & (0.0127) & (0.0184) \\
\vspace{-12pt} \\
\hspace{3mm}US-Born                      & 0.0262** & 0.0106** \\
                                         & (0.0114) & (0.0052) \\
\\
\textbf{\textit{Occupation}} \\
\hspace{3mm}Faculty                      & 0.0061 & -0.00208 \\
                                         & (0.0141) & (0.0177) \\
\vspace{-12pt} \\
\hspace{3mm}Postdoc/Grad                 & 0.0193 & 0.0432** \\
                                         & (0.0178) & (0.0177) \\
\vspace{-12pt} \\
\hspace{3mm}Other                        & 0.015 & 0.0114  \\
                                         & (0.0126) & (0.0083) \\
\vspace{-12pt} \\
\hline\hline
\end{tabular}





\begin{tablenotes}[flushleft]
\small
\item \hspace{-0.5em} This table shows, for single-R01 personnel, aggregated estimates of the average treatment effects (ATTs) of an interruption on being nonemployed in the US, split by whether or not a personnel is observed in the 2020 Decennial Census. They are obtained using our modified \cite{callaway2020dd} estimator. The effects are aggregated over the 5 years after the expiration of a lab’s grant. Standard errors are bootstrapped and clustered at the expiring-R01-level. *, **, and *** indicate statistical significance at the 10\%, 5\%, and 1\% levels.
\end{tablenotes}
\end{threeparttable}
\end{table}

Breaking out the results by place of birth, Table \ref{tab:tab-noadmin-census-subsample-maintext} shows that about 70\% of US-born personnel induced to nonemployment are found in the 2020 Decennial Census. 
In contrast, all of the foreign-born personnel induced to nonemployment leave the US.\footnote{Appendix Table \ref{tab:tab-censusonly} shows that single-R01 foreign-born personnel who experience a funding interruption are, if anything, more likely to leave the US than the point estimate for nonemployment suggests. Though interrupted US-born personnel are also less likely to be present in a Decennial Census, the effect is about half the size of their point estimate for nonemployment.} Thus, even though interruptions have a greater impact on the US-born's entrance into nonemployment, \emph{conditional} on being induced into nonemployment, the foreign-born are more likely to stay outside the US, likely reflecting visa employment requirements or attachment to their home countries (e.g.,~for family reasons).





Table \ref{tab:tab-noadmin-census-subsample-maintext} also breaks out the estimates by occupation, showing that trainees are most likely to leave the US. Among postdocs and graduate students, 70\% of the nonemployment effect is associated with being absent from the 2020 Decennial Census. In contrast, nearly all faculty and 60\% of personnel in other occupations that are induced to nonemployment by an interruption are present in the 2020 Decennial Census. The larger effect for trainees likely reflects the mobility of a relatively young population, who have less attachment to the US labor market and are more willing to find a job abroad, and are thus disproportionately pushed by interruptions out of the US scientific ecosystem.


\hypertarget{publications}{%
\subsection{Presence in Science: Publications}\label{publications}}

Although single-R01 personnel are more likely to leave the US after an interruption, possibly to the detriment of US science, the social welfare loss may be ameliorated if they are still working in science outside the US. We examine this possibility -- that personnel induced to leave the US by interruptions are still involved in scientific research -- by linking personnel to publications in the PubMed database, which allow us to track scientific activity beyond the US border. We then decompose the nonemployment effect by publishing activity, which is analogous to the exercise, in Section \ref{presence-in-the-united-states}, of decomposing the effect by presence/absence in the 2020 Decennial Census. 




To do this, we again create two new outcome variables. First, an indicator taking a value of one if a person is nonemployed in the US \emph{and} has at least one publication in a given year (zero otherwise). Second, an indicator taking a value of one if a person is nonemployed in the US \emph{and} has no publications in a given year (zero otherwise).


Table \ref{tab:pubnoadmin-tbl-subsample-main} displays our estimates for single-R01 labs. After an interruption, personnel are 0.4~pp more likely to be nonemployed and publishing and 2.6~pp more likely to be nonemployed and \emph{not} publishing. Thus, the vast majority -- 87\% ($2.6/(2.6+0.4)$) -- of interrupted single-R01 personnel that are pushed into nonemployment are also publishing less actively, suggesting that they are less likely to be participating in the scientific enterprise.  



\begin{table}[htbp]
\centering
\begin{threeparttable}
\caption{\label{tab:pubnoadmin-tbl-subsample-main}Effect of Interruptions on Publishing for Nonemployed Personnel from Single-R01 Labs}

\begin{tabular}{lcc} \hline\hline
& Nonemployed & Nonemployed   \\
& Publishing & Not Publishing   \\
\textit{Subample} & (1) & (2)  \\
\hline
\vspace{-12pt} \\
All Personnel                           & 0.004 & 0.026** \\
                                        & (0.006) & (0.009) \\
\\
\textbf{\textit{Place of Birth}} \\
\hspace{3mm}Foreign-Born                & -0.007 & 0.025 \\
                                        & (0.011) & (0.018) \\
\vspace{-12pt} \\
\hspace{3mm}US-Born                      & 0.011* & 0.025*** \\
                                         & (0.006) & (0.010) \\
\\
\textbf{\textit{Occupation}} \\
\hspace{3mm}Faculty                      & 0.010 & -0.004 \\
                                         & (0.014) & (0.012) \\
\vspace{-12pt} \\
\hspace{3mm}Postdoc/Grad                 & 0.003 & 0.059*** \\
                                         & (0.013) & (0.021) \\
\vspace{-12pt} \\
\hspace{3mm}Other                        & 0.007 & 0.020*  \\
                                         & (0.007) & (0.012) \\
\vspace{-12pt} \\
\hline\hline
\end{tabular}


\begin{tablenotes}[flushleft]
\small
\item \hspace{-0.25em}This table shows, for single-R01 personnel, aggregated estimates of the average treatment effects (ATTs) of an interruption on being nonemployed in the US, split by whether or not a personnel is observed with a publication in PubMed. They are obtained using our modified \cite{callaway2020dd} estimator. The effects are aggregated over the 5 years after the expiration of a lab’s grant. Standard errors are bootstrapped and clustered at the expiring-R01-level. *, **, and *** indicate statistical significance at the 10\%, 5\%, and 1\% levels.
\end{tablenotes}
\end{threeparttable}
\end{table}


Looking at the breakdown by place of birth, we see that nearly all of the foreign-born personnel pushed to nonemployment are less likely to publish. Meanwhile, some US-born personnel induced into nonemployment do remain active in science, with about 30\% continuing to publish at similar rates after an interruption.


The breakdown by occupation in the lab suggests that, once again, the results are most dramatic for non-faculty. About 96\% of graduate students and postdocs and 76\% of other lab personnel that become nonemployed after an interruption are also less likely to publish. In contrast, nearly all faculty induced to nonemployment continue to publish at a similar rate after an interruption.


Taken as a whole, the results of this section and Section \ref{presence-in-the-united-states} suggest that about half of single-R01 personnel who become nonemployed after an interruption leave the US and most (87\%) reduce their publication rates. Thus, not only do interruptions to single-R01 labs cause personnel to depart early from universities, but they also push personnel to leave the US and to reduce contributions to the scientific enterprise in the form of publications. The size of this effect for trainees (graduate students and postdocs) is particularly policy-relevant, suggesting that funding delays hurt the social return to investments in human capital.\footnote{In Appendix Table \ref{tab:pubnoadmincensus-tbl}, we further decompose the nonemployment effect for US-born and foreign-born personnel in single-R01 labs, by both presence in the 2020 Decennial Census and publishing. Among foreign-born personnel induced to nonemployment, nearly all are neither in the 2020 Decennial Census nor publishing. Among the US-born induced to nonemployment, the differences are less stark. About half remain in the US but are less likely to publish. 
}

\hypertarget{earnings}{%
\subsection{Earnings and Job Mobility}\label{earnings}}

So far, we have seen that interruptions prematurely push some single-R01 personnel out of universities, into nonemployment, out of the US, and out of science altogether. Losing highly-trained research personnel is likely detrimental to the scientific enterprise, but it is not clear how interruptions affect other career outcomes. Thus, in this section, we further examine the damage that funding delays can do to the careers of lab personnel by estimating the impact of interruptions on earnings and job switching behavior.



\hypertarget{wage-outcomes}{%
\subsubsection{Earnings}\label{wage-outcomes}}

The \textit{ex ante} effects of interruptions on earnings are ambiguous. If interruptions cause personnel to  hastily depart from their current jobs at universities, taking less well-fitting jobs and slowing their career progression through academia, then earnings may decrease relative to the counterfactual of belonging to a continuously-funded lab. For instance, if a post-doc scrambles to find another job, their advancement to a tenure-track faculty position may be delayed or derailed entirely, resulting in lower earnings. Alternatively, if interruptions spur personnel to get private sector jobs, their earnings may be higher than what they would have received by staying in academia.

\textbf{Event Studies.} Figure \ref{fig:es-wages}A displays event studies, for research personnel in both multiple- and single-R01 labs, using the arcsinh of total earnings as the outcome. As with employment outcomes, interruptions have no effect on the earnings of personnel in multiple-R01 labs. In contrast, personnel in single-R01 labs experience a sharp earnings decline that reaches almost 40\% after 2 years and about 60\% after 5 years.\footnote{To interpret effect sizes with an arcsinh outcome, we follow the suggestion in \cite{bellemare2020elasticities} that the small-sample bias correction in \cite{kennedy1981semilog} suffices for most applications when the untransformed variable is large enough (untransformed mean greater than 10). \label{foot:bellemare}} These estimates almost surely overstate the impact because the sample includes personnel with zero earnings \emph{in the US}. However, as suggested by our analysis of nonemployment and presence in the 2020 Decennial Census (Section \ref{presence-in-the-united-states}), many of these personnel probably receive positive earnings outside the US, which we cannot observe using our US-based administrative/tax data.

To obtain more plausible estimates of the impacts of interruptions on earnings, we identify a subset of personnel who are ``fully-attached'' to the US labor market, defined as having positive earnings in \emph{all} years -1 to 5. Figure \ref{fig:es-wages}B shows that, as with the full sample, personnel in multiple-R01 labs are mostly unaffected by an interruption. 
Personnel in single-R01 labs again experience a sharp earnings decline after an interruption -- 15\% by year 1 and 38\% by year 5. 
Thus, our earnings results are not completely driven by the exit of research personnel from the US labor market; even conditional on employment in the US, personnel from single-R01 labs experience substantial and permanent earnings declines after a funding interruption.

Conditioning on positive earnings \emph{after} treatment raises concerns about whether changes in sample composition bias our estimates. Our estimates overstate the earnings impact if, for example, interruptions cause personnel without US citizenship or permanent residency to select into the fully-attached sample by accepting lower-paying jobs to remain in the US (rather than taking higher paying jobs overseas). Our estimates will also be too large (in magnitude) if the earnings impact of interruptions is smaller for personnel who select \emph{out} of the fully-attached sample (e.g.,~if the jobs they take outside the US are higher paying than their counterfactual US jobs). On the other hand, our results may understate the earnings impact by excluding personnel who are induced by interruptions to either permanent (e.g.,~retirement) or temporary (e.g.,~to care for children) nonemployment, thus truly receiving zero earnings. Though we cannot definitively determine the direction of the bias, we suspect that omitting personnel with true nonemployment spells is likely to dominate other sources of selection into and out of the fully-attached sample, in which case our estimate is a lower bound on the true effect of interruptions on earnings.\footnote{In Appendix Section \ref{nonus-earnings-calc}, we perform a back-of-the-envelope calculation which suggests that, if the true earnings effect is zero, then then among non-fully attached personnel, interrupted personnel have to earn 1.2 log points (or over 200\%) more than continuously-funded personnel, which intuitively seems unlikely.}




\textbf{Aggregated Static Effects.}  Table \ref{tab:tab-earnings-subsample} displays the aggregate post-treatment effects for the full sample and the fully-attached subsample, broken out separately for personnel in multiple- and single-R01 labs. For both samples, the earnings of multiple-R01 personnel change minimally in response to an interruption. For single-R01 personnel, however, consistent with the event studies, there are large and statistically significant interruption effects, with those in the full sample and fully-attached samples experiencing post-interruption earnings declines of about 40\% and 20\% respectively.



\begin{table}[htbp]
\centering
\begin{threeparttable}
\caption{\label{tab:tab-earnings-subsample}Effect of Interruptions on Arcsinh Earnings}

\begin{tabular}{lcc} \hline\hline
& Arcsinh Earnings & Arcsinh Earnings  \\
& No Attachment Restriction & Fully-Attached  \\
\textit{Subample} & (1) & (2) \\
\hline
\vspace{-12pt} \\
\textbf{Panel A: Single-R01 Labs} \\
\\
All Personnel                           & -0.523*** & -0.227***  \\
                                        & (0.137) & (0.063) \\
\\
\textbf{\textit{Occupation}} \\
\hspace{3mm}Faculty                      & -0.138 & -0.047 \\
                                         & (0.156) & (0.089) \\
\vspace{-12pt} \\
\hspace{3mm}Postdoc/Grad                 & -0.743*** & -0.215* \\
                                         & (0.277) & (0.131) \\
\vspace{-12pt} \\
\hspace{3mm}Other                        & -0.498*** & -0.402** \\
                                         & (0.169) & (0.177) \\
\\
\textit{Personnel Count}                 & 4,200 & 3,400 \\
\vspace{-12pt} \\
\hline
\vspace{-12pt} \\
\textbf{Panel B: Multiple-R01 Labs} \\
\\
All Personnel                            & 0.0493 & -0.059* \\
                                         & (0.097) & (0.035) \\
\\
\textit{Personnel Count}                 & 13,500 & 11,000 \\
\vspace{-12pt} \\
\hline\hline
\end{tabular}





\begin{tablenotes}[flushleft]
\small
\item \hspace{-0.25em}This table shows aggregated estimates of the average treatment effects (ATTs) of an interruption on the arcsine of total earnings. They are obtained using our modified \cite{callaway2020dd} estimator. The effects are aggregated over the 5 years after the expiration of a lab's grant. Panel A shows the results for personnel in a single-R01 lab and Panel B shows the results for personel in multiple-R01 labs. Column (1) uses the full sample of all research personnel and column (2) uses the subsample that have positive earnings in all periods from one period before expiry to five years after expiry (i.e., the fully attached subsample). Standard errors are bootstrapped and clustered at the expiring-R01-level. *, **, and *** indicate statistical significance at the 10\%, 5\%, and 1\% levels.
\end{tablenotes}
\end{threeparttable}
\end{table}

Breaking the results out by occupation, we see that the effects are most pronounced for non-faculty, with trainees (postdocs and grad students) and other personnel (e.g., staff and undergrads) experiencing post-interruption earnings declines of 20\% and 34\%. In contrast, and in tandem with small employment effects, faculty experience modest (and statistically insignificant) declines of 5\%.\footnote{For faculty, the sample with no attachment restrictions may provide a better estimate of the earnings impact since nonemployed faculty are overwhelmingly likely to be physically present in the US in 2020 (see Table \ref{tab:tab-noadmin-census-subsample-maintext}) and thus observed nonemployment are likely to be true nonemployment spells rather than exits from the US.} Thus, not only do non-faculty bear the brunt of the displacement effects of interruptions, they also suffer the largest long-term decline in earnings.\footnote{The larger estimate for ``Other'' occupations may be due to its heterogeneous composition in terms of education levels and skills. The sample is more likely to consist of personnel (e.g.,~administrative staff, undergraduates) who have less advanced degrees or different skills than faculty or trainees, thus affecting their resiliency to employment shocks \citep{hoynes2012suffers, davis2021stem}.}

Overall, we interpret the decline in earnings as evidence that the effect of interruptions is not simply to shift the allocation of personnel across sectors and borders, but to also disrupt their careers.

\hypertarget{job-switching}{%
\subsubsection{Job Mobility}\label{job-switching}}

In addition to earnings, we assess the career stability of personnel in the wake of an interruption by examining their job switching behavior. In general, job switching can be an important mechanism through which workers' earnings recover from adverse labor market events (e.g.,~\cite{oreopoulos2012recession}). However, this may not be true in the context of research personnel who wish to remain in the academic sector post-interruption. For instance, a postdoc who switches to a new postdoc position will forgo earnings from more lucrative jobs \citep{cheng2023postdoc}, in which case the combination of lower earnings and higher job switching may be a sign of greater career instability.

\begin{table}
\centering
\begin{threeparttable}
\caption{\label{tab:tab-newein}Effect of Interruptions on Job Changes for Fully-Attached Personnel from Single-R01 Labs}

\begin{tabular}[t]{lccc}
\hline\hline
\vspace{-12pt} \\
& New University & \hspace{1cm} & New Non-University \\
& (1) & \hspace{1cm} & (2) \\
\midrule
& 0.0370**  && -0.0218 \\
& (0.0178) && (0.0189) \\
\vspace{-12pt} \\
\hline\hline
\end{tabular}

\begin{tablenotes}[flushleft]
\small
\item \hspace{-0.25em}This table shows, for single-R01 personnel, aggregated estimates of the average treatment effects (ATTs) of an interruption on the count of new university and non-university EINs that pay a personnel in a given year. They are obtained using our modified \cite{callaway2020dd} estimator. The effects are aggregated over the 5 years after the expiration of a lab’s grant. Standard errors are bootstrapped and clustered at the expiring-R01-level. *, **, and *** indicate statistical significance at the 10\%, 5\%, and 1\% levels.
\end{tablenotes}
\end{threeparttable}
\end{table}

To examine job switching behavior, we construct two variables that count the numbers of \emph{new} academic and industry jobs a personel has in a given year.\footnote{More precisely, we count the number of EINs from which a personnel received earnings in a given year but not the previous year. These variables are constructed as a count of new jobs rather than a binary variable indicating a new job for Census disclosure reasons. Since it is a count, the percentage point interpretation is approximate.} Table \ref{tab:tab-newein} shows difference-in-difference estimates, for single-R01 personnel in the fully-attached subsample, of an interruption's impact on these job count variables. After an interruption, personnel are 3.7~pp more likely to have a new university job in a given year and no more likely to move into a new industry job. Thus, to the extent that single-R01 personnel switch jobs after an interruption, they opt for a new (presumably less desirable) academic job. In combination with the earnings decline of the previous section, we interpret these job switching results as evidence of interrupted personnel being pushed onto a less stable academic career track rather than being pushed into higher-paying industry jobs.

\hypertarget{robustness_main}{%
\subsection{Robustness}\label{robustness_main}}

\hypertarget{robustness}{%
\subsubsection{Controlling for Resubmissions}\label{robustness}}




A natural concern about the credibility of our estimates is whether differences, across interrupted and continuously-funded labs, in project or PI quality, lead to violations of the parallel trend assumption. Overall, we find parallel trends quite plausible in our setting because raw means and event studies suggest that interrupted and continuously-funded personnel trend similarly prior to grant expiry (Section \ref{employment-outcomes}) and balance statistics suggest that they are similar across a variety of pre-treatment observables (Section \ref{analysis-sample}). Nevertheless, we probe the robustness of our main results by controlling for the perceived quality of an R01 renewal application -- specifically, the number of resubmissions an R01 renewal application went through before approval.\footnote{We are grateful to Ian Hutchins for the suggestion.}

\begin{table}
\centering
\begin{threeparttable}
\caption{\label{tab:resub-tab} Main Results, for Personnel from Single-R01 Labs, Controlling for Resubmissions}

\begin{tabular}{lcc}
\hline\hline
         &  & Controlling for \\
         &  No Controls           & Resubmissions \\
         &       (1)      & (2) \\
\midrule
\textit{Dep. Variable} \\
\vspace{-6pt} \\
Nonemployed       & 0.03024*** & 0.03204*** \\
                  & (0.01168) & (0.01068) \\
\vspace{-12pt} \\
US University     & -0.02104 & -0.02231 \\
                  & (0.01843) & (0.02024) \\
\vspace{-12pt} \\
US Industry       & -0.009197 & -0.009729 \\
                  & (0.0147) & (0.01796) \\
\vspace{-12pt} \\
asinh(Earnings)   & -0.2268*** & -0.1792*** \\
                  & (0.06271) & (0.06271) \\
\vspace{-12pt} \\
\hline\hline
\end{tabular}


\begin{tablenotes}[flushleft]
\small
\item \hspace{-0.25em}This table shows, for single-R01 personnel, aggregated estimates of the average treatment effects (ATTs) of an interruption on three mutually exclusive employment outcomes: 1) nonemployed in the US, 2) employed at a US university, and 3) employed in US industry. It also shows the ATT for the arcsinh of total earnings, which is estimated using subsample of research personnel that have positive earnings in all periods from one period before expiry to five years after expiry (i.e.,~ the fully-attached subsample).
They are obtained using our modified \cite{callaway2020dd} estimator. The effects are aggregated over the 5 years after the expiration of a lab’s grant. Column (1) reproduces the main estimates from Table \ref{tab:tab-main-subsample} and column (2) shows the same estimates, but controlling for the number of resubmissions. Standard errors are bootstrapped and clustered at the expiring-R01-level. *, **, and *** indicate statistical significance at the 10\%, 5\%, and 1\% levels.
\end{tablenotes}
\end{threeparttable}
\end{table}

In our context, a resubmission is an application for the renewal of R01 funding that follows an initial unsuccessful attempt. Thus, the number of resubmissions is a coarser version of the score that the NIH awards a renewal application.\footnote{In 2009, the NIH went from allowing two resubmissions to one resubmission (in rare cases, more resubmissions were possible under either regime).} Moreover, the need to resubmit an application can itself lead to a funding interruption -- indeed, we estimate that an additional resubmission is associated with a 12~pp increase in the likelihood of interruption. Thus, controlling for the number of resubmissions allows us estimate the effects of interruptions within groups of labs that have similar levels of perceived quality and similar probabilities of experiencing a resubmission-induced interruption.


Table \ref{tab:resub-tab} shows that controlling for the number of resubmissions has a negligible impact on our employment and earnings outcomes. Though we would undoubtedly prefer a finer measure of renewal application quality (such as confidential reviewer scores), these results suggest that the effects of interruptions are not primarily driven by unobservable project/PI quality differences.

\hypertarget{sample-split-by-interruption-time}{%
\subsubsection{Long and Short Funding Delays}\label{sample-split-by-interruption-time}}




Our analysis so far has not allowed treatment effects to vary by the duration of an interruption to a lab's R01 funding. However, longer interruptions may lead to larger effects if, for example, PIs can find short-term support (e.g.,~university-provided bridge funding) for personnel until R01 funding arrives. In this case, short delays can be weathered and the effects of interruptions would be concentrated in labs that experience longer funding delays. 


To examine whether interruption effects vary by delay length, we create two treated samples -- personnel in labs with ``short'' interruptions (delays of 30 to 90 days) and ``long'' interruptions (delays of 90 days or more).\footnote{90 days is approximately the median length of a funding interruption.} We then obtain estimates for each of these treatment groups, comparing them to our usual continuously-funded control group but omitting the other treatment group.\footnote{We opt for two separate treatment groups because the CS estimator does not accommodate continuous or discrete multi-valued treatments.}

Table \ref{tab:att-dosage} shows these estimates for our main employment and earnings outcomes. As with our baseline 30-day definition of an interruption, both short and long interruptions have imprecise aggregate effects on the probabilities of being at a university or in industry. For US nonemployment, the short interruptions have a larger impact (4.5~pp) than long interruptions (1.1~pp, statistically insignificant). Finally, for earnings, the impacts of short and long interruptions are quite similar at 25\% and 20\%, respectively. 

Therefore, both short and long funding delays impact the careers of single-R01 lab personnel, suggesting that, if alternative funding is available, its ability to mitigate the effects of interruptions on personnel does not vary by the length of the funding delay. Moreover, it appears that labs with particularly long delays (i.e.,~labs with particularly intense treatments) do not drive our results and that universities/PIs have trouble bridging even relatively short funding gaps.

\begin{table}[htbp]
\centering
\begin{threeparttable}
\caption{\label{tab:att-dosage}Effect of Interruptions by Length of Interruption}
\centering

\begin{tabular}[t]{lcccc}
\hline\hline
\vspace{-12pt} \\
 & Nonemployed  & US University & US Industry & asinh(Earnings) \\
 & (1)  & (2) & (3) & (4) \\
\midrule
\textit{Interruption Length} \\
\vspace{-6pt} \\
30-90 days & 0.0451*** & -0.0192 & -0.0259 & -0.253*** \\
           &  (0.0160) & (0.0232)  &  (0.0228) & (0.0884)\\
\addlinespace
>90 days & 0.0109 & -0.0234 & 0.0125 & -0.1952*** \\
         & (0.0149) & (0.0271) &  (0.0169) & (0.0684) \\
\addlinespace
\textit{Subsample} & All & All & All & Fully Attached \\
\vspace{-12pt} \\
\hline\hline
\end{tabular}

\begin{tablenotes}[flushleft]
\small
\item \hspace{-0.25em}This table shows, for single-R01 personnel, aggregated estimates of the average treatment effects (ATTs) of an interruption on three mutually exclusive employment outcomes: 1) nonemployed in the US, 2) employed at a US university, and 3) employed in US industry. It also shows the ATT for the arcsinh of total earnings, which is estimated using subsample of research personnel that have positive earnings in all periods from one period before expiry to five years after expiry (i.e.,~ the fully-attached subsample). They are obtained using our modified \cite{callaway2020dd} estimator. The first row contains estimates for labs with 'short' interruptions and the second row contains estimates for labs with 'long' interruptions.  The effects are aggregated over the 5 years after the expiration of a lab’s grant. Standard errors are bootstrapped and clustered at the expiring-R01-level. *, **, and *** indicate statistical significance at the 10\%, 5\%, and 1\% levels.
\end{tablenotes}
\end{threeparttable}
\end{table}

\hypertarget{conclusion}{%
\section{Conclusion}\label{conclusion}}

In this paper, we study how delays in grant funding, or interruptions, affect the careers of researchers -- not only the PIs who run the lab, but also the personnel they hire, including trainees (e.g., ~graduate students, post-docs) and staff (e.g., ~research scientists, lab managers). Using a combination of public-use NIH grant data and transaction-level data on grant expenditures, we identify the research labs to which personnel belong and measure lab-level variation in funding delays for the subset of R01 grants that are eventually \emph{successfully renewed}. By linking personnel to comprehensive US tax data, Decennial Censuses, and publications, we then track the evolution of their labor market outcomes after grant expiry.


Using a difference-in-differences design, we find that when the renewal of a PI's R01 is interrupted, their hired personnel are immediately and persistently less likely to work in the US at least up to five years later (the PIs themselves are unaffected). These effects are sizeable, amounting to about one-third of the nonemployment effect of childbearing for mothers with a Ph.D. in the biological sciences \citep{cheng2021childcare}. Most of the interruption effect is driven by personnel permanently leaving the US and reducing their publishing activity. Those who remain employed in the US also earn substantially less than their continuously-funded peers and seem to have less stable jobs, as evidence by their job switching activity. These results are concentrated among those less attached to universities -- that is, graduate students, postdocs, and lab staff -- illustrating the susceptibility to interruptions of personnel that are not on long-term, university-supported contracts.

Compared to foreign-born personnel, the US-born are more responsive to funding delays, being twice as likely to be induced into US nonemployment. This is perhaps counterintuitive, however, the foreign-born often face work or study requirements to remain in the US and barriers to reentry if they leave. These restrictions may lead the foreign-born to be less responsive to funding set-backs. Moreover, though interruptions are more likely to induce the US-born to nonemployment, we find that, \emph{conditional} on becoming nonemployed, the foreign-born are more likely to stay outside the US.


Among the constellation of policies and institutions affecting the labor market outcomes of the research workforce, how important are renewal delays for R01 grants?  Based on our estimates, funding delays account for about 5\% of nonemployment among research personnel within our sample five years post-interruption. This is comparable to the effect of green card delays on whether US doctorates stay in the US -- estimates from \cite{kahn2020impact} imply that being from a country affected by permanent residency visa caps (China or India) accounts for 7.4\% of departures from the US. Thus, eliminating funding delays could have a meaningful effect on the retention of scientific talent within the US. 


The fact that we find non-trivial effects arising from a relatively modest source of funding instability also suggests that larger sources of instability, which are difficult to study because they typically stem from aggregate shocks (e.g.,~the federal budgeting process) or otherwise impact everyone (e.g.,~general uncertainty in the grant system), likely have even larger impacts on the labor market outcomes of the research workforce. Thus, our results are informative not only about the effects of funding delays \textit{per se}, but also about 
the effects of funding instability more generally.

Our finding that the effects of interruptions are confined to personnel in labs supported by a single R01 further suggests that policies to more broadly allocate NIH grants across many labs may have the unintended consequence of also increasing funding instability for all labs \citep{reardon2007nih, russo2008should}. Even when application renewals are uncertain, labs can still engage in longer-term planning if they are also supported by other non-expiring grants, which becomes less likely if grants are spread more thinly across an increasing number of labs.

More fundamentally, our results raise questions about the wisdom of tying science funding to the yearly appropriations process. While some flexibility may be necessary to respond to short-term needs or opportunities (e.g.,~emergencies such as the COVID-19 pandemic or unexpected scientific breakthroughs), the long-term nature of science suggests that it may be prudent to appropriate some portion of science agencies' budgets for multiple years at a time. 


This leads to the question: what are candidate policy solutions for reducing or eliminating funding delays (or at least mitigating their impact)? The answer depends in part on the mechanisms driving the effects of funding delays. One possible mechanism is that PIs do not receive sufficient notice or information that enables them to find other temporary sources of funding (such as bridge funding from their institution). In this case, providing PIs with more and/or earlier information about their chances of renewal success would help to mitigate the effects of interruptions. On the other hand, it may simply be that such short-term funding options are scarce, such that even giving PIs earlier notice would be of minimal help. If so, policymakers might consider the provision of ``insurance'' or short-term liquidity to affected labs. Alternatively, a more fundamental policy change would be to increase the share of trainees funded by fellowships that are not tied to a specific PI.

Overall, our results contain two main lessons. First, operational details matter in science funding \citep{williams2023delays, duflo2017plumber}. The yearly delay in passing a federal budget and the possibility of funding interruptions are issues that scientists and institutions are keenly aware of and try to prepare for. As one PI noted, ``this period of gaps can be very stressful, in general it is hard to plan research in advance, when getting most of your funding in installments and uncertainty.'' Even within our sample of research-intensive universities, the impact of funding interruptions on personnel is not trivial. 

Second, not only are grants inputs into knowledge production, but also into the production of future scientific human capital \citep{jiang2024}. The benefits of policies and new funding models that reduce the occurrence of funding disruptions may be understated if we focus on research output (e.g.,~publication-derived outcomes) without considering the impact on the people working in these fields. Indeed, our data cannot tell us anything about the personal effects of these uncertainties on the lives and well-being of lab personnel. In addition, funding uncertainty and disruptions might shape the beliefs of those who have yet to start (or are in the early stages of) their scientific career, and therefore influence the composition of the future scientific labor force. More work shedding light on the interaction of funding certainty with expectations and career choice may be a fruitful line of future work \citep{ganguli2019postdocs, aucejo2020impact}. 

\bibliography{interrupted-employees}  

\clearpage

\appendix
\setcounter{figure}{0}
\renewcommand{\thefigure}{A\arabic{figure}}
\renewcommand*{\theHfigure}{\thefigure}
\setcounter{table}{0}
\renewcommand{\thetable}{A\arabic{table}}
\renewcommand*{\theHtable}{\thetable}
\numberwithin{equation}{section}

\hypertarget{appendix-figtab}{%
\section{Figures and Tables}\label{appendix-tables}} 

\begin{wrapfigure}{r}{1\textwidth}
\begin{center}
{\includegraphics[width=0.8785\textwidth]{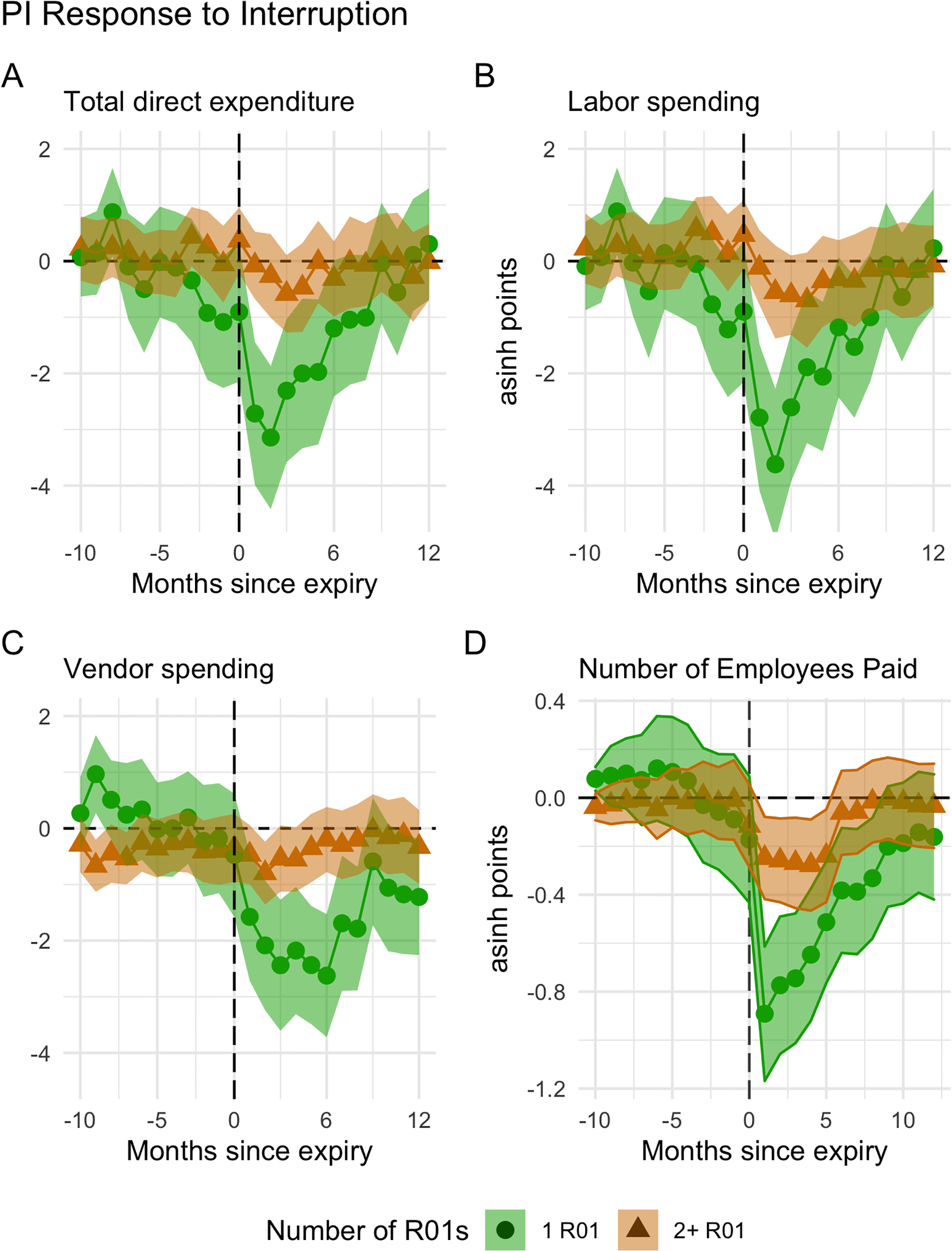}}
\end{center}
\begin{singlespace}
\caption{\small This figure shows the main results from \citet{tham2023interrupted}. The figure shows event-study estimates (with 95\% confidence intervals) of the difference in lab spending and number of employees between PIs of interrupted and uninterrupted R01s. Details are available in \citet{tham2023interrupted}.}\label{fig:lab_spending}
\end{singlespace}
\end{wrapfigure}

\clearpage



\begin{figure}[htbp]
TWFE With an Alternative Control Group

{\centering \includegraphics{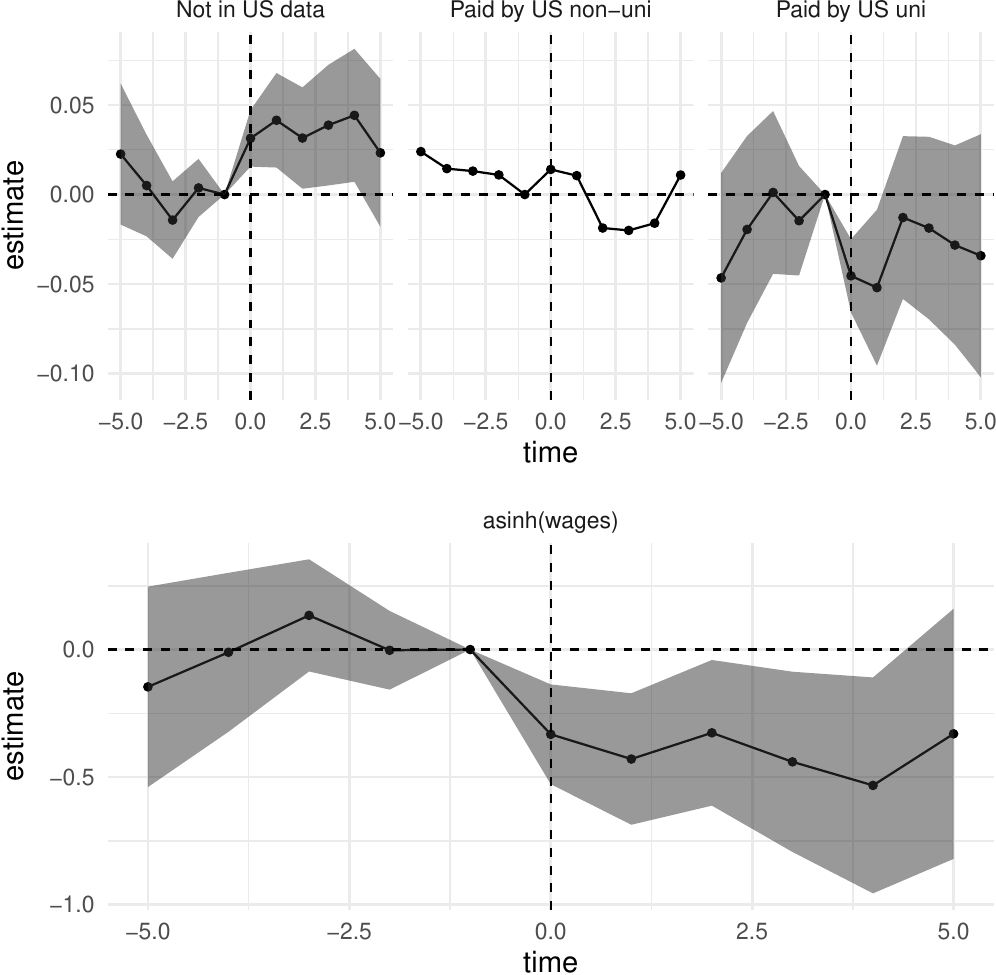} 
}
\caption{This figure shows the event studies using an alternative control group, employees in labs with interrupted R01s and multiple R01s. The first row shows event studies for the three mutually exclusive placement outcomes: absent from US data, paid by a US university, and paid by a US non-university. The second row shows the event study for arcsinh-transformed wages.}\label{fig:es-alt-control}
\end{figure}
\clearpage


\begin{table}
\centering
\begin{threeparttable}
\caption{\label{tab:tab-ownuniv}Effect of Interruptions on Being at Original University or Different University for Different Subsamples}
\centering
\begin{tabular}[t]{lcc}
\toprule
 & At Original &  \\
& (UMETRICS) & At New \\
Subsample \hspace{1cm} & University & University \\
\midrule
All & -0.0211  & 0.0240 \\
& (0.0215) & (0.0182) \\
\addlinespace
\midrule
\addlinespace
Foreign-born & 0.0177 & 0.00705 \\
& (0.0281) & (0.0224)\\
\addlinespace
US-born & -0.0390 & 0.0325 \\
& (0.0250) & (0.0228) \\
\addlinespace
\midrule
\addlinespace
Faculty & \makecell[c]{-0.0408**\\(0.0202)} & \makecell[c]{0.0110\\(0.0209)}\\
\addlinespace
Postdoc/Grad & \makecell[c]{-0.0775**\\(0.0331)} & \makecell[c]{0.0613\\(0.0393)}\\
\addlinespace
Others & \makecell[c]{0.0110\\(0.0359)} & \makecell[c]{0.00830\\(0.0188)}\\
\bottomrule
\end{tabular}
\begin{tablenotes} [flushleft]
\small
\item [] This table shows aggregated estimates of the average treatment effects (ATTs) of in our single-R01 sample of an interruption on the probability of being at a person's original UMETRICS university and of being at any other university. These indicate whether: (1) the personnel receives positive earnings from \emph{their own UMETRICS university} (i.e.,~the university employing them at the time of R01 expiry) and (2) the personnel receives positive earnings from an IPEDS university other than their own. The first two rows are for all single-R01 employees, while the next two are for foreign-born employees, and the last two for US-born employees.The effects are aggregated over the 5 years after the expiration of a lab’s grant. The estimates are obtained using a modified \cite{callaway2020dd} estimator. Standard errors are bootstrapped and clustered at the interrupted R01 level. *, **, and *** indicate statistical significance at the 10\%, 5\%, and 1\% levels.
\end{tablenotes}
\end{threeparttable}
\end{table}

\begin{table}[htbp]
\centering
\begin{threeparttable}
\hypertarget{tab:tab-censusonly}{\caption{\label{tab:tab-censusonly}Effect of Interruptions on Being Observed in Subsequent Decennial Censuses}}

\begin{tabular}[c]{lc}
\toprule
Subsample \hspace{4cm} & In Census \\
\midrule
All & -0.0294** \\
& (0.0149)\\
\addlinespace
\midrule
\addlinespace
Foreign-born & -0.0390 \\
& (0.0397) \\
\addlinespace
US-born & -0.0157 \\
& (0.0183)\\
\addlinespace
\midrule
\addlinespace
Faculty & 0.0467 \\
& (0.0317)\\
\addlinespace
Postdoc/Grad & -0.0234 \\
& (0.0393)\\
\addlinespace
Other & -0.0432* \\
& (0.0246)\\
\bottomrule
\end{tabular}
\begin{tablenotes}[flushleft]
\small
\item [] This table shows aggregated estimates of the average treatment effects (ATTs) of in our single-R01 sample of an interruption on the probability of being found in a Decennial Census (2000, 2010, and 2020). Note that this necessarily can only contain data for Decennial Census years.The first row is for all employees of single-R01 labs, while the following rows are for sub-samples of this group. The effects are aggregated over the 5 years after the expiration of a lab’s grant. The estimates are obtained using a modified \cite{callaway2020dd} estimator. Standard errors are bootstrapped and clustered at the interrupted R01 level. *, **, and *** indicate statistical significance at the 10\%, 5\%, and 1\% levels.
\end{tablenotes}

\end{threeparttable}
\end{table}

\begin{table}[htbp]
\centering
\begin{threeparttable}
\caption{\label{tab:pubnoadmincensus-tbl}Effect of Interruptions on Publishing and Presence in 2020 Decennial Census if Absent From Tax Data}
\centering

\begin{tabular}[c]{lcccc}
\toprule
Subsample & Not Publishing & Not Publishing & Publishing & Publishing \\
& Not In 2020 Census & In 2020 Census & Not In 2020 Census & In 2020 Census\\
\midrule
Foreign-born & 0.0357** & -0.00977 & -0.00774 & 0.000950 \\
& (0.0173) & (0.0116) & (0.0116) & (0.00631)\\
\addlinespace
\midrule
\addlinespace
US-born & 0.008823* & 0.01815** & 0.003393 & 0.009127 \\
& (0.005007) & (0.008856) & (0.004069) & (0.005595)\\
\bottomrule
\end{tabular}
\begin{tablenotes}[flushleft]
\small
\item [] This table shows aggregated estimates of the average treatment effects (ATTs) in our single-R01 sample of an interruption on the not being found in the administrative data, split by two things: whether or not a person is observed with a publication in MEDLINE, and whether or not a person is observed in the 2020 Decennial Census.  For each group, the coefficients should sum to the overall estimate of not finding people in the administrative data. The first four rows are for all foreign-born people in single-R01 labs. The second four are for US-born people in single-R01 labs.  The effects are aggregated over the 5 years after the expiration of a lab’s grant. The estimates are obtained using a modified \cite{callaway2020dd} estimator. Standard errors are bootstrapped and clustered at the interrupted R01 level. *, **, and *** indicate statistical significance at the 10\%, 5\%, and 1\% levels.
\end{tablenotes}
\end{threeparttable}
\end{table}

\begin{table}[htbp]
\centering
\begin{threeparttable}
\caption{\label{tab:eventstudies_sector} Event Studies for Sector of Employment}
\centering
\begin{tabular}[t]{lccc|ccc}
\toprule
 &  \multicolumn{3}{c}{Single R01} &  \multicolumn{3}{c}{Multiple R01s} \\
 & Non- & US & US  & Non- & US & US \\
  & -employed & University & Industry & -employed & University & Industry \\
\midrule
T-5	& 0.00322	& 0.00508	& -0.00830 & -0.00132	& 0.00772	& 0.00640 \\
	& (0.0153)	& (0.0120)	& & (0.00623)	& (0.00661) & \\
 \multicolumn{3}{c}{ \vspace{-8pt} } &  \\
T-4	& -0.00415	& 0.00767	& -0.00353 & -0.000122	& 0.00404	& 0.003916 \\
	& (0.0139)	& (0.0126)	& & (0.00606)	& (0.00654) & \\
 \multicolumn{3}{c}{ \vspace{-8pt} } &  \\
 T-3	& -0.0125	& 0.0110	& 0.00153 & 0.00243	& -0.00169	& 0.000746 \\
	& (0.0118)	& (0.0143)	& & (0.00630)	& (0.00757) \\
 \multicolumn{3}{c}{ \vspace{-8pt} } &  \\
 T-2	& 0.0101	& -0.0136	& 0.00353 & 0.000134	& 0.00496	& 0.00510 \\
	& (0.00921)	& (0.0127)	& & (0.00685)	& (0.00736) \\
  \multicolumn{3}{c}{ \vspace{-8pt} } &  \\
T-1	& -0.000914	& 0.00543	&-0.00451 & 0.00728	& -0.00996	& -0.00268 \\
	& (0.00954)	& (0.0212)	& & (0.00558)	& (0.00706) & \\
  \multicolumn{3}{c}{ \vspace{-8pt} } &  \\
T+0	& 0.0290***	& -0.0489***	& 0.0199 & -0.00330	& -0.00276	& -0.00606 \\
	& (0.0100)	& (0.0140)	& & (0.00536)	& (0.0103) & \\
  \multicolumn{3}{c}{ \vspace{-8pt} } &  \\
T+1	& 0.0313**	& -0.0396*	& 0.00831 & -0.00747	& 0.00184	& -0.00563 \\
	& (0.0125)	& (0.0205)	& & (0.00763)	& (0.0124) & \\
  \multicolumn{3}{c}{ \vspace{-8pt} } &  \\
T+2	& 0.0249*	& -0.00550	& -0.0194 & -0.00873	& 0.00534	& -0.00339 \\
	& (0.0139)	& (0.0239)	& & (0.00924)	& (0.0141) & \\
  \multicolumn{3}{c}{ \vspace{-8pt} } &  \\
T+3	& 0.0274	& -0.00398	& -0.0234 & -0.00714	& -0.00512	& -0.01226 \\
	& (0.0168)	& (0.0256)	& & (0.0122)	& (0.0183) & \\
  \multicolumn{3}{c}{ \vspace{-8pt} } &  \\
T+4	& 0.0379*	& 0.00232	& -0.0402 & -0.0248*	& 0.0187	& -0.00602 \\
	& (0.0203)	& (0.0303)	& & (0.0137)	& (0.0233) \\
  \multicolumn{3}{c}{ \vspace{-8pt} } &  \\
T+5	& 0.0342	& -0.00906	& -0.0251 & -0.0100	& 0.0276	& 0.0176 \\
	& (0.0239)	& (0.0351)	& & (0.0177)	& (0.0239) & \\
 
\bottomrule
\multicolumn{7}{p{0.9\linewidth}}{ \small  This table shows estimates underlying the event studies plotted in Figure \ref{fig:es-sector} of the effects of interruptions on three mutually exclusive employment outcomes: 1) nonemployed in the US , 2) employed at a US university, or 3) employed in US industry. They are obtained using our modified \cite{callaway2020dd} estimator. The expiration of a lab's grant takes place at year 0, and the estimated interruption effects range from 5 years before to 5 years after expiry. The left column is for personnel in a single-R01 lab and the right column is for personnel in a multiple-R01 lab. Standard errors are bootstrapped and clustered at the expiring-R01-level. *, **, and *** indicate statistical significance at the 10\%, 5\%, and 1\% levels. Standard errors not available for US Industry for Census disclosure avoidance reasons.}\\
\end{tabular}
\end{threeparttable}
\end{table}

\begin{table}[htbp]
\footnotesize
\centering
\begin{threeparttable}
\caption{\label{tab:rawmeans} Raw Means: Probability by Sector}
\centering
\begin{tabular}[t]{lccc|ccc}
\toprule
{\normalsize A} &   \multicolumn{6}{c}{ \small Nonemployed \hspace{1.25cm} } \\
&   \multicolumn{3}{c}{Single R01} & \multicolumn{3}{c}{Multiple R01s} \\
& Continuously &  &  & Continuously &  & \\
& Funded & Interrupted & Difference & Funded & Interrupted & Difference \\
\midrule

T-5	 & 	21.14\%	 & 	21.06\%	 & 	0.08	 & 	20.68\%	 & 	20.73\%	 & 	-0.05	 \\
T-4	 & 	16.19\%	 & 	15.74\%	 & 	0.45	 & 	17.25\%	 & 	16.61\%	 & 	0.64	 \\
T-3	 & 	11.59\%	 & 	10.30\%	 & 	1.29	 & 	13.59\%	 & 	13.11\%	 & 	0.48	 \\
T-2	 & 	8.42\%	 & 	7.50\%	 & 	0.92	 & 	9.30\%	 & 	8.63\%	 & 	0.67	 \\
T-1	 & 	4.19\%	 & 	3.41\%	 & 	0.78	 & 	5.07\%	 & 	4.95\%	 & 	0.12	 \\
T+0	 & 	4.13\%	 & 	5.80\%	 & 	-1.67	 & 	5.01\%	 & 	4.54\%	 & 	0.47	 \\
T+1	 & 	7.05\%	 & 	9.08\%	 & 	-2.03	 & 	8.44\%	 & 	7.37\%	 & 	1.07	 \\
T+2	 & 	9.34\%	 & 	10.65\%	 & 	-1.31	 & 	10.53\%	 & 	10.12\%	 & 	0.41	 \\
T+3	 & 	10.80\%	 & 	12.21\%	 & 	-1.41	 & 	12.31\%	 & 	12.15\%	 & 	0.16	 \\
T+4	 & 	11.95\%	 & 	13.67\%	 & 	-1.72	 & 	13.99\%	 & 	12.35\%	 & 	1.64	 \\
T+5	 & 	12.26\%	 & 	13.77\%	 & 	-1.51	 & 	15.62\%	 & 	14.50\%	 & 	1.12	 \\

\midrule
\multicolumn{7}{c}{ \vspace{-8pt} } \\
{\normalsize B} &   \multicolumn{6}{c}{ \small US University \hspace{1.25cm} } \\
&   \multicolumn{3}{c}{Single R01} & \multicolumn{3}{c}{Multiple R01s} \\
& Continuously &  &  & Continuously &  & \\
& Funded & Interrupted & Difference & Funded & Interrupted & Difference \\
\midrule

T-5	 & 	62.91\%	 & 	64.04\%	 & 	-1.13	 & 	66.51\%	 & 	65.30\%	 & 	1.21	 \\
T-4	 & 	69.22\%	 & 	72.10\%	 & 	-2.88	 & 	71.26\%	 & 	71.20\%	 & 	0.06	 \\
T-3	 & 	76.02\%	 & 	79.11\%	 & 	-3.09	 & 	76.90\%	 & 	76.83\%	 & 	0.07	 \\
T-2	 & 	82.90\%	 & 	85.58\%	 & 	-2.68	 & 	83.88\%	 & 	84.22\%	 & 	-0.34	 \\
T-1	 & 	91.01\%	 & 	93.92\%	 & 	-2.91	 & 	91.37\%	 & 	90.92\%	 & 	0.45	 \\
T+0	 & 	90.77\%	 & 	88.67\%	 & 	2.10	 & 	90.53\%	 & 	90.07\%	 & 	0.46	 \\
T+1	 & 	81.58\%	 & 	80.46\%	 & 	1.12	 & 	80.59\%	 & 	81.67\%	 & 	-1.08	 \\
T+2	 & 	74.75\%	 & 	76.88\%	 & 	-2.13	 & 	73.28\%	 & 	74.07\%	 & 	-0.79	 \\
T+3	 & 	68.73\%	 & 	71.37\%	 & 	-2.64	 & 	67.69\%	 & 	67.96\%	 & 	-0.27	 \\
T+4	 & 	63.53\%	 & 	67.09\%	 & 	-3.56	 & 	62.77\%	 & 	65.26\%	 & 	-2.49	 \\
T+5	 & 	60.23\%	 & 	63.02\%	 & 	-2.79	 & 	57.50\%	 & 	61.77\%	 & 	-4.27	 \\

\midrule
\multicolumn{7}{c}{ \vspace{-8pt} } \\
{\normalsize C} &   \multicolumn{6}{c}{\small US Industry \hspace{1.25cm} } \\
&   \multicolumn{3}{c}{Single R01} & \multicolumn{3}{c}{Multiple R01s} \\
& Continuously &  &  & Continuously &  & \\
& Funded & Interrupted & Difference & Funded & Interrupted & Difference \\
\midrule

T-5	&	15.95\%	&	14.90\%	&	1.05	&	12.81\%	&	13.97\%	&	-1.16	 \\
T-4	&	14.59\%	&	12.16\%	&	2.43	&	11.49\%	&	12.19\%	&	-0.70	 \\
T-3	&	12.39\%	&	10.59\%	&	1.80	&	9.51\%	&	10.06\%	&	-0.55	 \\
T-2	&	8.68\%	&	6.92\%	&	1.76	&	6.82\%	&	7.15\%	&	-0.33	 \\
T-1	&	4.80\%	&	2.67\%	&	2.13	&	3.56\%	&	4.13\%	&	-0.57	 \\
T+0	&	5.10\%	&	5.53\%	&	-0.43	&	4.46\%	&	5.39\%	&	-0.93	 \\
T+1	&	11.37\%	&	10.46\%	&	0.91	&	10.97\%	&	10.96\%	&	0.01	 \\
T+2	&	15.91\%	&	12.47\%	&	3.44	&	16.19\%	&	15.81\%	&	0.38	 \\
T+3	&	20.47\%	&	16.42\%	&	4.05	&	20.00\%	&	19.89\%	&	0.11	 \\
T+4	&	24.52\%	&	19.24\%	&	5.28	&	23.24\%	&	22.39\%	&	0.85	 \\
T+5	&	27.51\%	&	23.21\%	&	4.30	&	26.88\%	&	23.73\%	&	3.15	 \\

\bottomrule
\multicolumn{7}{p{0.87\linewidth}}{ \scriptsize This table shows the number that are ploted in plotted in Figure \ref{fig:fraction-by-sector}: the average probability that a personnel is in one of three mutually exclusive employment categories: 1) nonemployed in the US (Panel A), 2) employed at a US university (Panel B), or 3) employed in US industry (Panel C). These probabilities are calculated from five years before R01 grant expiration to five years after. Numbers in the left column are for personnel in a single-R01 lab and numbers in the right column are for personnel in a multiple-R01 lab. Both groups are split into continuously-funded and interrupted labs.}\\
\end{tabular}
\end{threeparttable}
\end{table}

\begin{table}[htbp]
\centering
\begin{threeparttable}
\caption{\label{tab:eventstudies_earnings}Event Studies for Earnings}
\centering
\begin{tabular}[t]{lcc|cc}
\toprule
 &  \multicolumn{2}{c}{  All Research Personnel } &  \multicolumn{2}{c}{ Fully-Attached Subsample  } \\
   \multicolumn{2}{c}{ \vspace{-8pt} } &  \\
&  Single R01 &  Multiple R01s &  Single R01 &  Multiple R01s \\
\midrule
T-5	&	-0.116	&	-0.00241	&	-0.133	&	0.0356	 \\
	&	(0.119)	&	(0.0544)	&	(0.122)	&	(0.0579)	 \\
  \multicolumn{2}{c}{ \vspace{-8pt} } &  \\
T-4	&	0.0273	&	0.0287	&	-0.0113	&	0.0286	 \\
	&	(0.125)	&	(0.0550)	&	(0.118)	&	(0.0587)	 \\
  \multicolumn{2}{c}{ \vspace{-8pt} } &  \\
T-3	&	0.0774	&	-0.00755	&	0.00774	&	-0.0383	 \\
	&	(0.110)	&	(0.0605)	&	(0.109)	&	(0.0598)	 \\
  \multicolumn{2}{c}{ \vspace{-8pt} } &  \\
T-2	&	-0.00196	&	0.00758	&	-0.0298	&	0.0545	 \\
	&	(0.0891)	&	(0.0638)	&	(0.0857)	&	(0.0625)	 \\
  \multicolumn{2}{c}{ \vspace{-8pt} } &  \\
T-1	&	-0.0205	&	-0.0431	&	-0.0377	&	0.0358	 \\
	&	(0.0896)	&	(0.0562)	&	(0.0832)	&	(0.0579)	 \\
  \multicolumn{2}{c}{ \vspace{-8pt} } &  \\
T+0	&	-0.321***	&	0.0365	&	-0.0506	&	-0.00859	 \\
	&	(0.124)	&	(0.0609)	&	(0.0633)	&	(0.0212)	 \\
  \multicolumn{2}{c}{ \vspace{-8pt} } &  \\
T+1	&	-0.456***	&	0.0785	&	-0.163***	&	-0.0171	 \\
	&	(0.140)	&	(0.0873)	&	(0.0422)	&	(0.0271)	 \\
  \multicolumn{2}{c}{ \vspace{-8pt} } &  \\
T+2	&	-0.446***	&	0.0404	&	-0.216***	&	-0.0668	 \\
	&	(0.156)	&	(0.108)	&	(0.0646)	&	(0.0489)	 \\
   \multicolumn{2}{c}{ \vspace{-8pt} } &  \\
T+3	&	-0.549***	&	0.0368	&	-0.292***	&	-0.0418	 \\
	&	(0.192)	&	(0.130)	&	(0.0892)	&	(0.0426)	 \\
   \multicolumn{2}{c}{ \vspace{-8pt} } &  \\
T+4	&	-0.774***	&	0.156	&	-0.375***	&	-0.115*	 \\
	&	(0.232)	&	(0.162)	&	(0.104)	&	(0.0659)	 \\
   \multicolumn{2}{c}{ \vspace{-8pt} } &  \\
T+5	&	-0.832***	&	-0.0692	&	-0.462***	&	-0.186**	 \\
	&	(0.294)	&	(0.214)	&	(0.129)	&	(0.0861)	 \\
\bottomrule
\multicolumn{5}{p{0.7\linewidth}}{ \small  This table shows estimates underlying the event studies plotted in Figure \ref{fig:es-wages} of the effects of interruptions on the arcsinh of total earnings. They are obtained using our modified \cite{callaway2020dd} estimator. The expiration of a lab's grant takes place at time 0, and the estimated interruption effects range from 5 years before to 5 years after expiry. The first and third column show estimates for personnel in a single-R01 lab and the second and fourth column show estimates for personnel in a multiple-R01 lab. The left side shows estimates for the full sample of all research personnel and the right side shows estimates for the subsample of research personnel that have positive earnings in all periods from one period before expiry to five years after expiry (i.e., the fully attached subsample). Standard errors are bootstrapped and clustered at the expiring-R01-level. *, **, and *** indicate statistical significance at the 10\%, 5\%, and 1\% levels.}\\
\end{tabular}
\end{threeparttable}
\end{table}

\begin{table}[htbp]
\centering
\begin{threeparttable}
\caption{\label{tab:eventstudies_multicontrol} Event Studies for TWFE With an Alternative Control Group}
\centering
\begin{tabular}[t]{lcccc}
\toprule
 & Nonemployed & US University & US Industry & Earnings \\
 \midrule
T-5	&	0.0226	&	-0.04657	&	0.02397	&	-0.1463	 \\
	&	(0.02009)	&	(0.02991)	&		&	(0.2005)	 \\
   \multicolumn{2}{c}{ \vspace{-8pt} } &  \\
T-4	&	0.005031	&	-0.0195	&	0.014469	&	-0.01077	 \\
	&	(0.01448)	&	(0.02663)	&		&	(0.159)	 \\
    \multicolumn{2}{c}{ \vspace{-8pt} } &  \\
T-3	&	-0.01425	&	0.001164	&	0.013086	&	0.1339	 \\
	&	(0.01104)	&	(0.0232)	&		&	(0.1122)	 \\
    \multicolumn{2}{c}{ \vspace{-8pt} } &  \\
T-2	&	0.003689	&	-0.01464	&	0.010951	&	-0.002881	 \\
	&	(0.008275)	&	(0.0156)	&		&	(0.0787)	 \\
    \multicolumn{2}{c}{ \vspace{-8pt} } &  \\
T-1	&		&		&		&		 \\
	&		&		&		&		 \\
    \multicolumn{2}{c}{ \vspace{-8pt} } &  \\
T+0	&	0.03138***	&	-0.0454***	&	0.01402	&	-0.3323***	 \\
	&	(0.008057)	&	(0.01045)	&		&	(0.09978)	 \\
    \multicolumn{2}{c}{ \vspace{-8pt} } &  \\
T+1	&	0.04146***	&	-0.052**	&	0.01054	&	-0.4293***	 \\
	&	(0.01346)	&	(0.02221)	&		&	(0.1315)	 \\
    \multicolumn{2}{c}{ \vspace{-8pt} } &  \\
T+2	&	0.03154**	&	-0.01285	&	-0.01869	&	-0.3265**	 \\
	&	(0.01444)	&	(0.02319)	&		&	(0.1457)	 \\
    \multicolumn{2}{c}{ \vspace{-8pt} } &  \\
T+3	&	0.03876**	&	-0.01872	&	-0.02004	&	-0.4403**	 \\
	&	(0.01719)	&	(0.026)	&		&	(0.1803)	 \\
    \multicolumn{2}{c}{ \vspace{-8pt} } &  \\
T+4	&	0.04424**	&	-0.02822	&	-0.01602	&	-0.5327**	 \\
	&	(0.01894)	&	(0.02838)	&		&	(0.216)	 \\
    \multicolumn{2}{c}{ \vspace{-8pt} } &  \\
T+5	&	0.02328	&	-0.03418	&	0.0109	&	-0.3305	 \\
	&	(0.02107)	&	(0.03464)	&		&	(0.2502)	 \\
 
\bottomrule
\multicolumn{5}{p{0.7\linewidth}}{ \small  This table shows estimates underlying the event studies plotted \ref{fig:es-alt-control} in which use an alternative control group, employees in labs with interrupted R01s and multiple R01s. There are event studies for the three mutually exclusive placement outcomes: absent from US data, paid by a US university, and paid by a US non-university. The last column shows the event study for arcsinh-transformed wages. Standard errors are clustered at the expiring-R01-level *, **, and *** indicate statistical significance at the 10\%, 5\%, and 1\% levels.}\\
\end{tabular}
\end{threeparttable}
\end{table}

\clearpage



\hypertarget{alternative-control-group}{%
\section{Alternative Control Group}\label{alternative-control-group}}

One concern about our identification strategy is whether there are unobserved differences between interrupted and uninterrupted labs that both cause interruptions and affect labor market outcomes of lab personnel irrespective of interruptions. In the main estimates we see that the interruptions have no effects among the multiple R01 group, despite interruptions indicating the same potential confounders as in the single R01 group. To visualize this, we use the same stacked-sample setup as above, but with a two-way fixed effects estimator, where we treat personnel in labs that also had interrupted R01s but had multiple R01s as the control group. Figure \ref{fig:es-alt-control} shows the event studies for this robustness test. The treatment dynamics for this alternative estimator and control group are similar to those in our main results, which suggests that our results are not driven by the estimator or primarily by interruption related confounders.


%

\hypertarget{interviews-with-principal-investigators-of-interrupted-r01s}{%
\section{Interviews with PIs of Interrupted R01s}\label{interviews-with-principal-investigators-of-interrupted-r01s}}

We conducted interviews with the Principal Investigators (PIs) of interrupted R01s to better understand how they perceive and respond to the threat of interruptions. We used the ExPORTER database \protect\hyperlink{appendix-exporter}{ExPORTER database} to identify interrupted R01s from 2019 to 2021  and then looked up the emails of those PIs on the \href{https://reporter.nih.gov/}{RePORTER} website.\footnote{RePORTER is an online tool for searching NIH grants data.} We first contacted PIs at universities where we are alumni (Boston University, Harvard University, and The Ohio State University), and then randomly selected a group to contact from all other universities. We successfully reached six PIs in total. Three of the PIs had multiple R01s at the time of interruption, and three of the PIs had one R01 at the time of interruption. These interviews were conducted between December 2022 and February 2023.

\clearpage
\newpage

\hypertarget{estimation-appendix}{%
\section{Estimation}\label{estimation-appendix}}
\addcontentsline{toc}{section}{Estimation}

\hypertarget{estimating-stacked-data-with-cs-estimator-callaway2020dd}{%
\subsection{Estimating stacked data with CS estimator}\label{estimating-stacked-data-with-cs-estimator-callaway2020dd}}
\addcontentsline{toc}{subsection}{Estimating stacked data with CS estimator}

Stacked data is usually estimated by OLS with cohort-specific unit and time fixed effects.\footnote{See the appendices of \cite{cengiz2019minwage} for an implementation and \cite{baker2021dd} for a discussion.} We instaed use the estimator developed by \cite{callaway2020dd} (``CS estimator'') because it has some desirable features such as more transparent weighting in the aggregation of group-time treatment effects, simultaneous confidence intervals, and a doubly robust modeling option. However, the CS estimator, as currently implemented via the Stata \textit{csdid} command \citep{csdid-package}, does not straightforwardly accommodate the case where controls have a well-defined ``treatment'' date. The rest of this section describes how we implement the CS estimator in this context.\footnote{This was originally described in the following \href{https://twitter.com/wytham88/status/1368963248335257600?s=20\&t=GfGbNt6iI3xODc5DfZfciQ}{tweet}.}

The main issue is that the \texttt{csdid} package treats all control units as being control units for all treatment cohorts, whereas in our case, control units belong to specific treatment cohorts. This means that if implemented without any modifications, for any given treatment cohort, the \texttt{csdid} package will use observations of control units from another treatment cohort in estimation. Consider, for instance, two treatment cohorts with treatments in 2000 and 2001 and units in each cohort spanning one year before and one year after (so the units in cohort 2000 span 1999 to 2001). The \texttt{csdid} package will compare treatment and control units belonging to the cohort 2001. However, control units from the year 2000 will also have observations in the year 2001 that \texttt{csdid} will use in estimation.

\begin{center}\includegraphics{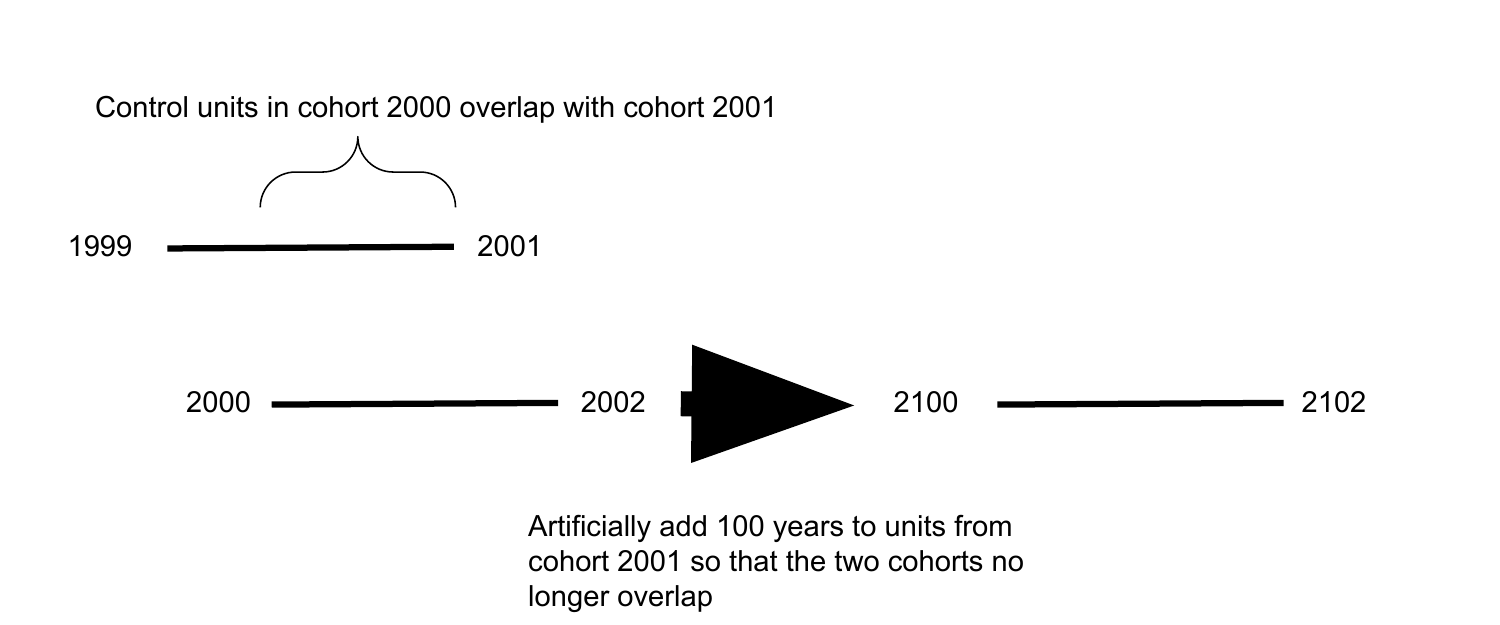} \end{center}

The goal is ensure that treated units within cohort are only compared to control units in the same cohort. This can be done by adding a large number to the calendar years for units in each cohort, thus artificially ``separating'' them from other cohorts. For instance, consider the example of two cohorts, 2000 and 2001, with the 2000 cohort spanning years 1999 to 2001 and the 2001 cohort spanning years 2000 to 2002. If we artificially add 100 years to the 2001 cohort, then it spans the years 2100 to 2102, which no longer overlaps with the 2000 cohort. Thus, the \texttt{csdid} package will no longer use observations from the 2000 cohort in 2001 for estimating effects in the 2001 cohort.

One drawback of this workaround is that it is no longer possible (or at least not straightforward) to aggregate effects by calendar years, since they have been relabelled with fake calendar years.

\hypertarget{nonus-earnings-calc}{%
\subsection{Implied non-US earnings}\label{nonus-earnings-calc}}
\addcontentsline{toc}{subsection}{Implied non-US earnings}

In Section \ref{wage-outcomes}, we estimate the difference in earnings between interrupted and continuously-funded personnel who are fully attached to the US labor market. This effect could be an overestimate if, for instance, treated, fully-attached personnel select into in to lower-paying jobs in the US (relative to the control group). Here we outline a calculation to understand the extent of this selection needed to to substantially alter our conclusions. For simplicity, we abstract from the difference-in-differences setting and assume a difference-in-means setup. 

Let $\beta$ be the true effect, $\hat{\beta}$ is the estimate that we report using the fully attached sample (approx. $-0.2$ log points). Let $a$ denote that an  individuals is fully attached to the US labor market and $b$ denotes otherwise. Thus, $Y_{1}^{a}$ and  $Y_{0}^{a}$ are the average log wages of the interrupted and continuously-funded personnel who are in the fully attached sample respectively; $n_{1}^{b}$ and $n_{0}^{b}$ are the proportions of non-fully attached personnel within the treated and control groups respectively. Similar definitions follow for group $b$.

$\beta$ and $\hat{\beta}$ are calculated as follows: 

\begin{align*}
    \beta &= Y_{1} - Y_{0} \\&= (1 - n_{1}^{b}) Y_{1}^{a} + n_{1}^{b} Y_{0}^{b} + (1 - n_{0}^{b}) Y_{0}^{a} + n_{1}^{b} Y_{0}^{b} \\
    \hat{\beta} &= Y_{1}^{a}  - Y_{0}^{a} 
\end{align*}

From these definitions it follows that, if $n_1^b = n_0^b = n^b$ \footnote{We make this assumption because the counts separated by treatment and control group have not passed disclosure review.}

\begin{align*}
    \beta - \hat{\beta} &= n_1^b (Y_1^b - Y_1^a) - n_0^b (Y_0^b - Y_0^a) \\
    (Y_1^b - Y_0^b) - (Y_1^a - Y_0^a) &= \frac{\beta - \hat{\beta}}{n_b}   
\end{align*}

That is, the bias scaled by the proportion of non-fully attached personnel tells us how large the treated-control wage differential has to be for the fully attached and non-fully attached samples. In our sample, $n_b = 0.2$, $\hat{\beta} = -0.2$. Under these assumptions, if the true $\beta = 0$, then the wage difference for the non-fully attached sample has to be 1 log point greater than the wage differential for the fully attached sample.

\hypertarget{kahn2020calc}{%
\subsection{Kahn and MacGarvie 2020 calculation}\label{kahn2020calc}}
\addcontentsline{toc}{subsection}{Kahn and MacGarvie 2020 calculation}

To calculate the proportion of departures from the US attributable to green card delays, we use the estimate from Table 1 Column (2) of and sample statistics from the \emph{N} and \emph{Stay Rate} rows of Appendix Table A1a in \cite{kahn2020impact} to calculate (1) the number of departures due to green card delays ($0.076 \times (N_{China} + N_{India})$) and (2) the total number of departures ($N_{RoW} * DepartRate_{RoW} + N_{India} * DepartRate_{India} + N_{China} * DepartRate_{China}$), where the departure rate is $1  - \text{Stay Rate}$.  The proportion of delays due to green card delays is (1) divided by (2).






\clearpage
\newpage

\hypertarget{data-appendix}{%
\section{Data Appendix}\label{data-appendix}}
\addcontentsline{toc}{section}{Data Appendix}

\hypertarget{appendix-census}{%
\subsection{IRS and Census Data}\label{appendix-census}}
\addcontentsline{toc}{subsection}{IRS and Census Data}

\textbf{W-2 tax records.} Form W-2 is an Internal Revenue Service (IRS) form that US employers must file listing the wages paid to an employee and taxes withheld. Each W-2 record contains an employee's tax identification number paired with the federal tax identification number (EIN) of an employer, and information on yearly wages. EINs allow us to identify which employers are universities (see IPEDS section below).

\textbf{Longitudinal Employer-Household Dynamics (LEHD).} LEHD data contain Unemployment Insurance (UI) wage records which track earnings and employment at a quarterly frequency. Since UI programs are administered at the state level, each record is an employee paired with the state tax identification number (SEIN) of the employer. However, the federal EIN is also available for most employee-SEIN pairs. Note that student stipends/personnel are not subject to unemployment insurance, and thus this income is not observed in the LEHD (though this income would appear on the W-2 forms).

\textbf{1040 Schedule C (1040-C) tax records.} The 1040-C tax records are available through the Integrated Longitudinal Business Database (ILBD) at the Census Bureau, and contain the population of all nonemployer firms in the United States \citep{goetz2021recent, jarmin2007integrated}. These 1040-C records capture earnings from self-employment.

\textbf{IPEDS.} We link the W-2 and LEHD data to a public-use list of university EINs from the Integrated Postsecondary Education Data System (IPEDS), so we can determine whether an individual is paid by a university.\footnote{The public-use list of university EINs from IPEDS can be found \href{https://nces.ed.gov/ipeds/datacenter/DataFiles.aspx}{here} under the title ``Directory Information''. We combine the datasets from 2002 to 2018.} IPEDS contains EINs for most U.S.-based universities, and all UMETRICS universities are in IPEDS.

\textbf{Decennial Censuses.}  These data help us distinguish between people who stop working but are still present in the US (no earnings, but present in the Census) and people who leave the US altogether (no earnings, and not present in the Census), allowing us to assess the extent to which interruptions cause members of the scientific workforce to leave the US.

The majority (about 67\%) of housing units self-responded to the 2000, 2010, and 2020 Decennial Censuses, that is, they replied to prompts to answer questions about who was living in their housing unit by mailing back a form or by phone (or in 2020 by filling out a web based form). Data is also collected from those who administer group quarters (e.g., colleges, assisted living facilities, prisons), from non-response followup operations where enumerators visit housing units, and though a number of other operations such as enumeration at transitory locations designed to count hard to count palpitations (including in 2020, use of administrative records if they were deemed high-quality after an attempt at non-response follow up).

\hypertarget{umetrics-herd}{%
\subsection{UMETRICS}\label{umetrics-herd}}
\addcontentsline{toc}{subsection}{UMETRICS Universities}

\begin{figure}
{\centering \includegraphics{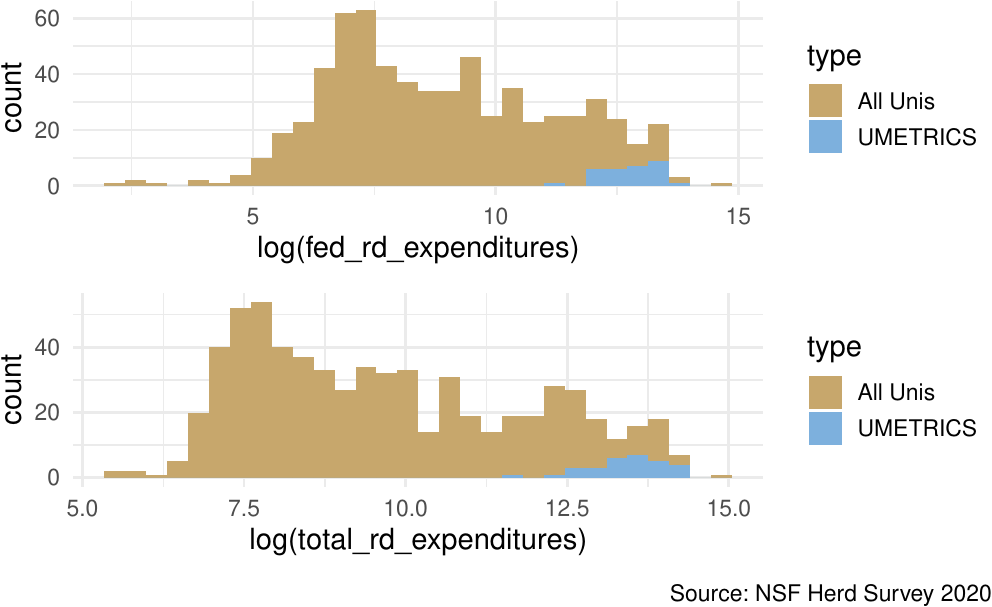} 
}
\caption{Histogram of the logarithm of total federal R\&D expenditures for all universities (including UMETRICS universities) in the NSF HERD survey and for UMETRICS universities.}\label{fig:umetrics-fedspend}
\end{figure}

UMETRICS is a database of administrative transaction-level data on payments made from university research grants to personnel and vendors. It is housed at the Institute for Research on Innovation and Science (IRIS) at the University of Michigan and is derived from university human resources records, sponsored projects, and procurement systems made available by participating universities. We use the 2020 release of UMETRICS, which contains data from 33 universities representing about one-third of US federal research expenditures \citep{umetrics2019review}. UMETRICS universities are research-intensive -- all are classified as R1 (Doctoral Universities -- Very high research activity) according to the Carnegie Classification System and rank in the top 20\% and top 25\% of universities by federal R\&D spending and total R\&D spending respectively (Figure \ref{fig:umetrics-fedspend}).

\hypertarget{appendix-exporter}{%
\subsection{ExPORTER}\label{appendix-exporter}}
\addcontentsline{toc}{subsection}{ExPORTER}

ExPORTER is publicly available data provided by the NIH.\footnote{\url{https://exporter.nih.gov/}} Multiple categories of data are available on ExPORTER: Projects, project abstracts, publications, link tables from projects to publications, patents, and clinical studies. We use the Projects data to calculate how long it takes for projects to be renewed, which in turn we use to define whether or not they were interrupted .

\hypertarget{defining-project-periods}{%
\subsubsection{Defining Project Periods}\label{defining-project-periods}}
\addcontentsline{toc}{subsubsection}{Defining project periods}

NIH projects are assigned a \textbf{core project number} that is used over multiple \textbf{project periods}. A project period is what we would conventionally call a ``grant'': some amount of funding guaranteed over a few years. At the end of each project period, the Principal Investigator (PI) of the project can apply to renew funding for that project. If the renewal application is successful, that begins a new project period. This interval between when a project period ends and when a new project period begins (after successful renewal) is the focus of this project. However, ExPORTER does not provide explicit identifiers for project periods so we have to use the other information provided in ExPORTER to determine when a project period started or ended.

The funds for a project period are allocated from the NIH to the project over multiple \textbf{budget periods}.\footnote{This is laid out in more detail in \href{https://grants.nih.gov/grants/policy/nihgps/html5/section_5/5.3_funding.htm}{Section 5.3 of the NIH Grants Policy Statement}.} Each budget period is recorded as a row in the ExPorter \emph{Projects} data.

For example, project number \emph{R01GM049850}, led by PI Jeffrey A. Simon, was funded from FY 1996 to FY 2017, except for FY 2013. Table \ref{tab:r01ex} below shows the records from its first eight years of funding. Each year the project was funded appears as a new row in the data. In the first year, the project was funded as a new project (application type 1), but then for each of the next three years was funded as a ``continuation'' (application type 5). The project is then funded as a ``renewal'' (application type 2) in FY 2000, then again as a ``continuation'' the next three years. Thus, we can infer that FY 1996 to FY 1999 constituted one project period. After that, the project had to be renewed, resulting in a new project period from FY 2000 to 2003.

The exact steps we use to determine project periods are:

\begin{enumerate}
\def\labelenumi{\arabic{enumi}.}
\tightlist
\item
  Indicate first budget period of a new project period if application type is 1, 2, or 9. Define the start date of the project period as the start date of the budget period.
\item
  Arrange budget periods by budget start date. Assign budgets that start after the first budget of a project period (as indicated by application type) to that project period, until the first budget of a new project period is reached.
\item
  Assign the project period end date as the latest budget end date of all budget periods assigned to the project period.
\end{enumerate}

\begin{table}

\caption{\label{tab:r01ex}Example of NIH ExPorter data before aggregation into project periods}
\centering

\begin{tabular}[t]{l|l|r|r|l}
\hline
PI Name & Core Project Num & Fiscal Year & Application Type & Comment\\
\hline
Simon, Jeffrey A & R01GM049850 & 1996 & 1 & New\\
\hline
Simon, Jeffrey A & R01GM049850 & 1997 & 5 & Continuation\\
\hline
Simon, Jeffrey A & R01GM049850 & 1998 & 5 & Continuation\\
\hline
Simon, Jeffrey A & R01GM049850 & 1999 & 5 & Continuation\\
\hline
Simon, Jeffrey A & R01GM049850 & 2000 & 2 & Renewal\\
\hline
Simon, Jeffrey A & R01GM049850 & 2001 & 5 & Continuation\\
\hline
Simon, Jeffrey A & R01GM049850 & 2002 & 5 & Continuation\\
\hline
Simon, Jeffrey A & R01GM049850 & 2003 & 5 & Continuation\\
\hline
\end{tabular}
\end{table}

\hypertarget{calculating-time-to-renewal}{%
\subsubsection{Calculating Time to Renewal}\label{calculating-time-to-renewal}}
\addcontentsline{toc}{subsubsection}{Calculating time to renewal}

Our treatment variable is the time between consecutive project periods for a given R01. For each pair of consecutive project periods, we calculate this as the number of calendar days between the expiration date of the earlier project period and the renewal date of the later project period. In cases where the expiration date is after the renewal date, we redefine the expiration date to be one calendar day before the renewal date. We also use this adjusted expiration date as the reference date for defining the periods of time we use to link employees to PIs or to link PIs to their other grants (described below).

\hypertarget{linking-pi-ids-to-project-periods}{%
\subsubsection{Linking PI IDs to project periods}\label{linking-pi-ids-to-project-periods}}
\addcontentsline{toc}{subsubsection}{Linking PI IDs to project periods}

PI IDs in ExPORTER are assigned at the row/budget period level. We assign a PI ID to a project period if a PI was assigned to any of the budget periods that constitute the project period.

\hypertarget{sample-construction}{%
\subsection{Sample Construction}\label{sample-construction}}
\addcontentsline{toc}{subsection}{Sample Construction}

An ideal dataset would allow us identify employees who were part of a PI's lab/research group that went through an R01 renewal. However, UMETRICS allows us to infer those relationships based on which employees a PI was paying around the time of renewal. The overall steps to construct our sample involve decisions at each of the following levels of data:

\begin{enumerate}
\def\labelenumi{\arabic{enumi}.}
\tightlist
\item
  R01
\item
  PI-R01
\item
  PI-R01-employee
\end{enumerate}

First, we find all pairs of expiring-renewed R01 project period pairs that were also successfully linked to UMETRICS. We keep all expiring-renewed pairs that were renewed within the same fiscal year to ensure that any observed delays in renewal were not due to unusual circumstances or data errors. This is also consistent with the NIH-level counterfactual we have in mind where the NIH funds the same projects within the same fiscal year without delay.

Next, we link the project periods in each expiring-renewed project period pair to their PI IDs. We retain all units where the PI ID appeared in both the expiring and renewed project periods. This leaves us with a set of (PI, expiring R01 project period, renewed R01 project period) triples. For simplicity, we refer to these as PI-expiring-R01 units.

Our next step is to link PI-expiring-R01 units to employees. For each PI-expiring-R01, we first fix a 12-month window that ends in the month the R01 was expiring. For example, if the expiring month is Dec 2021, the window is from Jan 2021 to Dec 2021. We then link each PI to all their NIH grants at in that time window based on the overlap between the 12-month window and the start and end dates of any project periods associated with the PI. This gives us a PI's portfolio of NIH grants in the 12-month period prior to expiry.

The next step is to find employees who were part of a PI's lab by finding employees who were paid any of the grants in this portfolio during the 12-month window. We first link the PI's grant portfolio to a crosswalk between NIH core project numbers and UMETRICS award numbers, an identifier in UMETRICS that accompanies each transaction. Through the award numbers, we then link to the UMETRICS employee dataset to obtain all employee numbers paid through the awards in the 12-month window.

\hypertarget{counting-r01-equivalents}{%
\subsection{Counting R01-Equivalents}\label{counting-r01-equivalents}}
\addcontentsline{toc}{subsection}{Counting R01-equivalents}

To take into account that PIs with more grants may have a buffer, we measure the size of a PI's grant portfolio based on the number of R01s they had around the time of R01 expiry. The process of constructing this measure is the same as the one for linking PIs to employees described in the previous section, except that we find all grants within a 24-month window that begins 11 months before and ends 12 months after the expiry date of the focal R01.

We include grants after expiry to allow for the possibility that PIs anticipating receiving more grants may be able or more willing to find ways to continue funding affected employees. This also assumes that the number of R01-equivalents is not affected by interruptions (i.e.,~not a post-treatment variable), which we think is reasonable in this context given the time lag between applying for and receving an R01.\footnote{E.g., A guide by NIAID suggests it can take 8 to 20 months upon applying \url{https://www.niaid.nih.gov/grants-contracts/timelines-illustrated}}

\hypertarget{defining-r01-equivalents}{%
\subsubsection{Defining R01-Equivalents}\label{defining-r01-equivalents}}
\addcontentsline{toc}{subsubsection}{Defining R01-equivalents}

Given the outsized importance of the R01, we use the number of R01s and R01-equivalents as our measure of a PI's grant portfolio. R01-equivalents are defined at the time of writing (2021) as ``activity codes DP1, DP2, DP5, R01, R37, R56, RF1, RL1, U01 and R35 from select NIGMS and NHGRI program announcements''.\footnote{\url{https://web.archive.org/web/20211221215217/https://grants.nih.gov/grants/glossary.htm}} However, the definition of R01-equivalent definitions can change slightly over time. We use the \href{https://web.archive.org/}{Internet Wayback Machine} to find R01-equivalent definitions going as far back as possible (late 2017) and include all activity codes ever defined as an R01. We also include all R35 grants rather than only those from NIGMS or NHGRI, as specified in the definition, as the R35 seems to be used similarly across the NIH (to provide long-term support for outstanding investigators e.g.,~see \href{https://web.archive.org/web/20211027025938/https://grants.nih.gov/grants/funding/ac_search_results.htm}{here}).

\end{document}